\documentclass[twocolumn]{article}
\usepackage{graphicx} 
\usepackage{titlesec}
\usepackage{amssymb}
\usepackage{amsmath}
\usepackage{multicol}
\usepackage{caption}
\usepackage{fancyhdr}
\usepackage{subfig}
\usepackage{placeins}

\usepackage[backend=biber,sorting=none]{biblatex}

\fancyheadoffset[RO,EL]{0pt}
\usepackage{siunitx}
\usepackage[]{csquotes}
\usepackage{url}

\usepackage{geometry}
\usepackage[backend=biber,sorting=none]{biblatex}
\title{Investigation of flow patterns in two-phase carbon dioxide in horizontal and vertical pipes}
\author{Alexander Siegfanz, Wolfgang Wagner and Uwe Janoske\\
{\itshape\footnotesize University of Wuppertal, Gau{\ss}stra{\ss}e 20, 42119 Wuppertal, Germany}}
\date{17 December 2023}

\titleformat*{\section}{\normalsize\bfseries}
\titleformat*{\subsection}{\normalsize\itshape}
\begin{document}

\maketitle

\begin{abstract}
\small
Modern particle detectors crucially depend on efficient cooling systems. Two-phase carbon dioxide
(CO$_2$) is a suitable solution as a cooling agent. This publication presents the observations and results
of investigations of horizontal and vertical flow of two-phase CO$_2$ at a temperature of
$T=-15\,^\circ$C and a pressure of approximately $23\,$bar.
Heat fluxes between $98.5\,$kW/m$^2$ and $200\,$kW/m$^2$ were applied to
the CO$_2$, covering the range expected to occur in the future ATLAS Pixel detector
being built for the high-luminosity phase of the Large Hadron Collider.
Flow speeds ranged from $11.8\,$m/s to $28.1\,$m/s.
Dedicated sensors measured the temperature and the pressure before and
after heating the CO$_2$.
Two-phase flow patterns occuring in the pipe after heating the CO$_2$ were recorded with a
high-speed camera. 
Stratified, wavy and slug flow are found to be the predominant patterns for horizontal flow,
while upward vertical flow is mainly found to be slug or churn.
Based on the recorded images the void fraction of the CO$_2$ after heating is determined and
compared for the different setups. The
results are summarised in a flow-pattern map. A clear distinction between vertical and
horizontal flow is found, with horizontal flow exhibiting a significantly higher void
fraction than vertical flow. Based on the pressure measurements, the pressure drop after
heating the CO$_2$ is measured and the corresponding Euler number is computed.
While the pressure drop increases with the heat flux for horizontal flow,
the pressure drop reduces with the heat flux in the case of upward vertical flow.
\end{abstract}

\section{Introduction}
Modern particle detectors in high-energy physics make heavy use of silicon-based detectors
for precisely measuring the trajectories of charged particles. The silicon sensors feature
a \(p\)-\(n\) junction which is reverse-biased to deplete the substrate of free charge carriers.
Electron-hole pairs created by ionisation are separated by the electric field in the sensor.
The sensors are connected to front-end electronics which amplifies, shapes, and digitizes the charge signal.
The sensor and the front-end chips form the key elements of a detector module which is the smallest
functional unit of a particle detector. 

The main source of the heat load of a detector module is the front-end electronics.
To avoid thermal runaway of the leakage currents in the sensors due to their temperature dependence
the detector modules must be cooled. Cooling the modules also keeps the
electronics at an acceptable temperature.
Typically, cooling  loops serve a series of multiple detector modules.
To keep multiple scattering of charged particles at a low level, the material budget related to the
cooling system must be kept at the necessary minimum.
At the same time, the operation conditions of the modules are required to be relatively homogenous,
and thus the temperature drop from the beginning to the end of a cooling loop must be small.
One common solution to meet these requirements is the application of two-phase cooling.
This is also the approach taken by the upgrade of the ATLAS Pixel detector for the high-luminosity
phase of the Large Hadron Collider (LHC) at the European Organisation for Nuclear Research (CERN). 
The current tracking detectors will be replaced to cope with the increased particle flux,
requiring a finer sensor granularity and higher radiation hardness~\cite{ATLAS:2016prj}.

A heat-flux of \SI{0.7}{\watt\per\cm^2}, which is mainly driven by the front-end readout chips,
is estimated for the pixel detector modules~\cite{ATLAS:2017svb}.
Considering the most critical operation mode of the modules with four front-end chips (quad modules)
and including a safety factor of 50\%, the maximum thermal power to be removed amounts to \SI{18.9}{\watt}
per module. In the outer barrel region of the future ATLAS pixel detector thermal contact to the 
cooling pipe is made through a base block of aluminium graphite with a contact length of \SI{20}{\milli\metre}
and a contact area of
\SI{91.1}{\milli\metre^2}, resulting in a maximum heat flux at the surface of the cooling pipe of approximately
\SI{200}{kW\per\metre^2}. 
The number of modules attached to one cooling loop varies between 16 and 36.
Carbon dioxide (CO$_2$) is chosen as a cooling agent. 
The baseline design of the cooling system requires a minimum mass-flow rate of \SI{4}{\gram\per\second}
to avoid dryout inside the cooling loops.
However, recent studies indicate that a mass-flow rate of \SI{2}{\gram\per\second} could be sufficient
to avoid dryout at the beginning of detector operation~\cite{Barroca:2703341}.
Motivated by this result, the presented study is focussed on the lower values of mass-flow rates,
namely \SI{1.5}{\gram\per\second} and \SI{2}{\gram\per\second}. 

Previous investigations of flow-pattern maps of two-phase CO$_2$ deal with horizontal flow in
microchannels~\cite{pettersen:2004,doi:10.1080/0145763049028010}. 
These two studies look at flow patterns in horizontal glass pipes with an inner diameter of
\SI{0.98}{\milli\metre} and a heat flux ranging from \SI{5}{kW\per\metre^2} to \SI{20}{kW\per\metre^2}.
The temperature range is \SI{0}{\degreeCelsius} to \SI{25}{\degreeCelsius}.
Yun and Kim developed flow-regime maps for a narrow rectangular channel and visualise two-phase
flow patterns for several conditions~\cite{YUN20041259}.
Gasche developed a setup to investigate the evaporation of CO$_2$ inside a microchannel with a
hydraulic diameter of \SI{0.8}{\milli\metre} and a saturation temperature of \SI{23.3}{\degreeCelsius}
for a heat flux of \SI{1.8}{kW\per\metre^2}~\cite{gasche2006}.
Plug, slug and annular flow were observed in this study.
Cheng and collaborators provide a summary of results concerning two-phase flow of CO$_2$ in microchannels and
macrochannels~\cite{cheng2008}.
A broad investigation of new prediction methods updates the flow-pattern maps for CO$_2$
on the basis of previous work and considers a wide range of conditions, including tube diameters
from $0.6$ to \SI{10}{\milli\metre}, mass velocities from $50$ to
\SI{1500}{\kilogram\per(\metre^2\second)} and heat fluxes from $1.8$ to
\SI{46}{kW\per\metre^2}~\cite{CHENG2008111}.  
A most recent study in the context of the ATLAS experiment gives a first overview of
flow patterns within the needed range, but is restricted to the vertical (upward and downward) 
flow~\cite{SCHMID2022110526}.

The work presented in this paper extends the investigation to horizontal flow of two-phase CO$_2$,
which is most relevant for cooling pipes of detector modules in the central part of the detector.
The gas phase results from heat transfer to the pipes.
Vertical flow was also measured in the same parameter space, allowing for a direct comparison
of horizontal and vertical flow patterns and properties.
The investigations consider heat fluxes in the range between \SI{100}{kW\per\metre^2} 
and \SI{200}{kW\per\metre^2} at a temperature of \(T=\SI{-15}{\degreeCelsius}\),
covering the expected operating range of the CO$_2$ cooling of the future ATLAS Inner Tracker
for the high luminosity LHC.
The measurements performed complement previous investigations
and provide further knowledge on CO$_2$ flow-pattern maps.

\section{Experimental approach and prototype}

\subsection{The experimental setup}
\begin{figure*}[!t]
  \centering
  \includegraphics[width=0.7\linewidth]{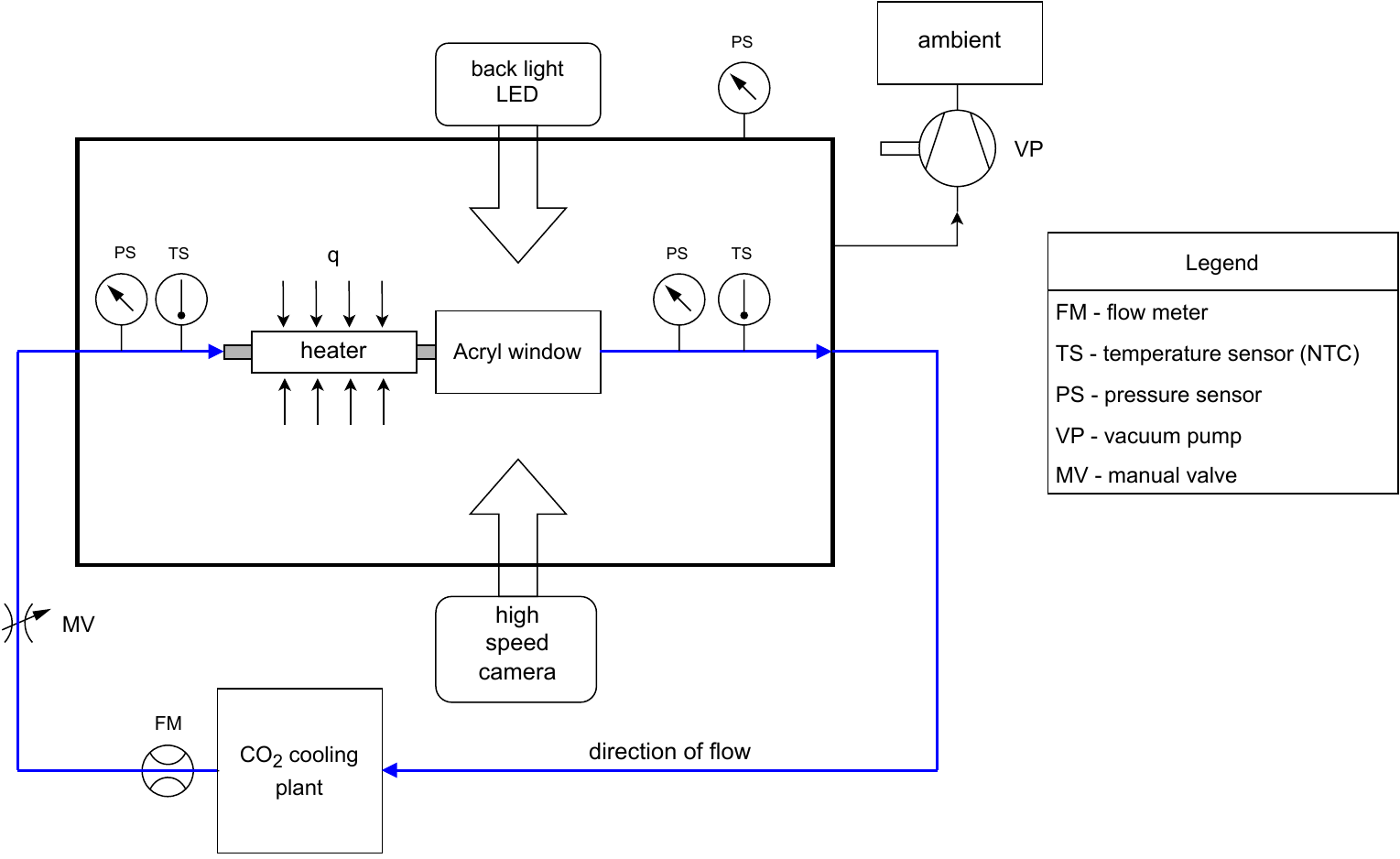}
  \caption{Schematic sketch of the experimental setup used for the measurements.
    The cooling plant delivers subcooled fluid CO$_2$ at a temperature of
    \(T=\SI{-15}{\degreeCelsius}\) and a pressure of approximately \SI{23}{\bar}.
    A homogeneous heat flux is applied to the pipe by means of a cartridge heater.
    The flow patterns induced by the formation of a gaseous phase are observed with a
    high-speed camera. Dedicated sensors measure the pressure and
    the temperature before and after applying the heat. The experimental setup is
    located in a vacuum chamber.
  }
  \label{fig:schematic}
\end{figure*}

\begin{figure}[!hb]
  \centering
  \includegraphics[width=\linewidth]{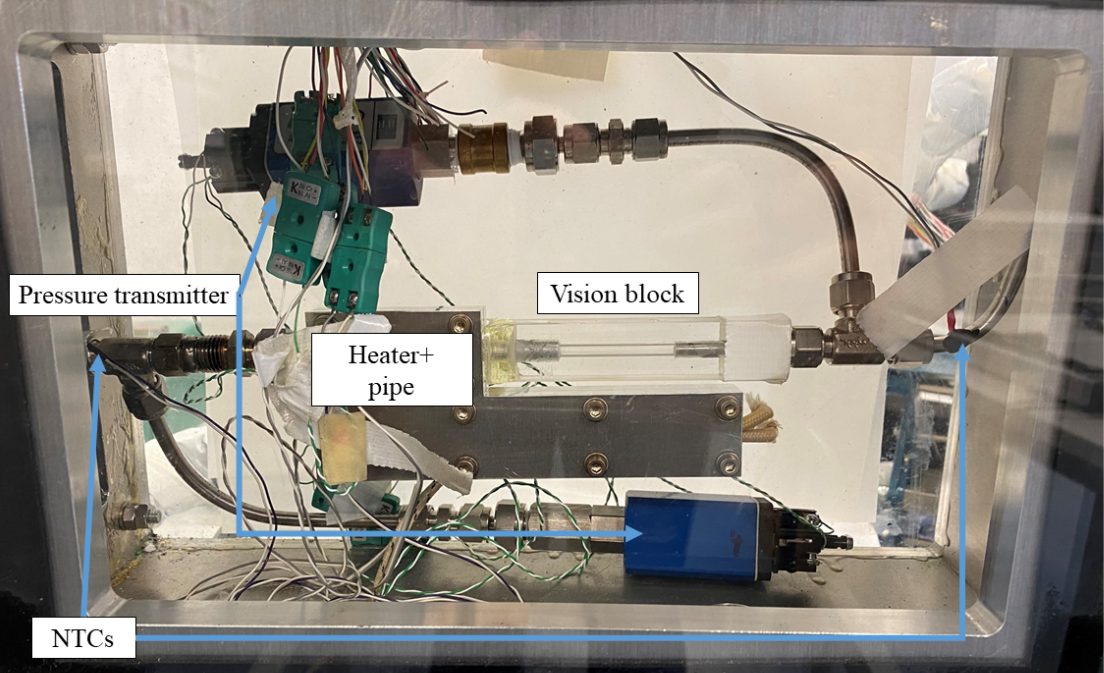}
  \caption{Photograph of one of the prototypes under investigation installed in the
    vacuum chamber and connected to pressure and temperature sensors.} 
  \label{fig:photo_setup}
\end{figure}

The test setup is engineered and constructed to test two-phase flow properties of CO$_2$
for pipes with inner diameters of \SI{3}{\milli\metre} and \SI{4}{\milli\metre}.
The azimuth angle \(\varphi\) specifies the orientation of the tube in the setup.
The setup facilitates the investigation of two-phase flow of CO$_2$ in horizontal
(\(\varphi=\ang{0}\)) and upward (\(\varphi=+\ang{90}\)) direction.
Figure~\ref{fig:schematic} shows a schematic sketch of the experimental setup used
for the measurements.
The cooling plant MARTA (Monoblock Approach for a Refrigeration Technical Application)
delivers the subcooled fluid CO$_2$ at a temperature of
\(T=\SI{-15}{\degreeCelsius}\) and a pressure of approximately \SI{23}{\bar}.
MARTA is an evaporating CO$_2$ cooling system designed for scientific research
purposes~\cite{marta} and is based on a method
called I-2PACL (Integrated 2-Phase Accumulator Controlled Loop).
The mass-flow rate is adjusted through a flow valve and the saturation temperature
can be controlled. Lower temperatures than \(T=\SI{-15}{\degreeCelsius}\) are not achievable.
A homogeneous heat flux is applied to the pipe by means of a cartridge heater with a
diameter of \SI{10}{\milli\metre}.
The flow patterns induced by the formation of a gaseous phase are observed with a
high-speed camera. Dedicated sensors measure the pressure and
the temperature before and after applying the heat.
Figure~\ref{fig:photo_setup} shows a photograph of one of the prototypes installed
in the vacuum chamber and connected to pressure and temperature sensors.

The measurement range of the pressure sensor
(DRTR-AL-10V-R100B) ranges from \SI{0}{\mega\pascal} to \SI{10}{\mega\pascal}. 
The uncertainty of the linearity for the pressure sensors is $\pm 0.2\%$. 
As shown in Figure~\ref{fig:photo_setup}, one sensor is located right before the
prototype and the second one directly behind the prototype.
Resistors with a negative temperature coefficient (NTC) are used as temperature sensors.
The NTCs have a resistance of \SI{100}{\kilo\ohm} with an uncertainty of \(\pm1\%\).
The sensors are glued directly on top of the pipe with the thermal conductive glue
LOCTITE STYCAST 2850FT using CAT9. The NTCs were calibrated with a PT100 and
a glycol bath, resulting in a residual uncertainty in the temperature measurement of \(\SI{0.1}{\kelvin}\). 
To apply the heat flux to the pipe a cartridge heater element with a thermal power of
\SI{150}{W} was used, the absolute precision of the set power being \(\pm10\%\). 
The readout of the pressure sensors and temperature sensors is realized with
an ELMB (Embedded Local Monitor Board)~\cite{elmb} and a LabVIEW program. 

A clamp mount was specifically designed and manufactured to keep the heater in place 
and to apply a homogeneous heat flux along the printed pipe. The resulting heated
length of the pipe was \SI{50}{\milli\metre}. The interfaces between heater and clamp 
and between clamp and pipe were covered with conductive paste.
The whole setup is assembled inside an aluminium vacuum box which is constantly pumped
out by a vacuum pump. 
The entire test setup can be used to investigate horizontal flow (\(\varphi=\ang{0}\)) 
and can be easily
rotated to measure flow in the upward direction (\(\varphi=+\ang{90}\)).

For visualisation of the occuring flow patterns a high-speed camera, IDT OS Model 3,
is used with an LED backlight.
The camera has a resolution of \(1280\times1024\) pixels and a variable focal length
between \(24\) and \SI{120}{\milli\metre}.
For each measurement the camera recorded \num{5000} pictures with a speed of \num{3000}
frames per second, thus covering a time period of \SI{1.666}{\second}. 

\subsection{The prototype}
\begin{figure}[!th]
  \centering
  \includegraphics[width=\linewidth]{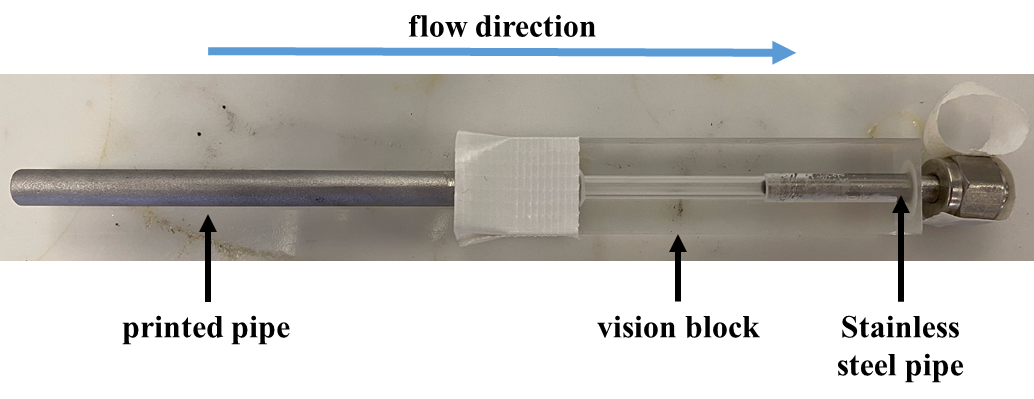}
  \caption{Photograph of the investigated prototype.}
  \label{fig:prototype}
\end{figure}

The investigated prototype consists of three main parts and is depicted in Figure~\ref{fig:prototype}. 
The first part on the left hand side of the picture is the printed pipe,
representing the cooling pipe used in a particle detector.
The prototype pipes were 3D printed from AlSi10Mg~\cite{print3D} 
with the selective laser melting (SLM) technique.
This procedure uses a high-power-density laser to melt metallic powder together.
The pipe had to be reworked with a reamer to achieve the correct inner diameters of
\SI{3}{\milli\metre} and \SI{4}{\milli\metre}.
The printed pipes were tested for pressure stability before they were added to the setup.
The second part of the prototype is the vision block made from acryl.
The block was drilled very precisely to avoid drill marks and thus achieve a clear
visibility of the carbon dioxide. 
The block was reamed to the corresponding inner diameter of the pipe. 
The third part is a stainless-steel pipe which serves as a connector to
the remaining setup. 
All three parts were glued end-to-end with an epoxy glue. Swagelok connectors
were used to connect the prototypes to the setup. 
The pressure stability of the whole prototype was tested up to 180 bar with nitrogen. 

\renewcommand\arraystretch{1.4}
\subsection{Measurement parameters}
The measurements were performed with four different geometrical pipe setups.
Pipes with an inner diameter of \(d=\SI{3}{\milli\metre}\) and
\(d=\SI{4}{\milli\metre}\) were used. The orientation of the pipes was chosen
to be horizontal (\(\varphi=\ang{0}\)) or upward (\(\varphi=+\ang{90}\)).
In addition, two mass-flow rates were considered for each of the geometrical setups:
\(\dot{m}=\SI{1.5}{\gram\per\second}\) and
\(\dot{m}=\SI{2.0}{\gram\per\second}\). 
The maximum mass-flow rate was given by the limitations of the MARTA cooling device.
The two diameters and two mass-flow rates lead to four different flow speeds \(v\) 
and mass velocities \(j\), as given in Table~\ref{tab:velocities}.
\begin{table}[t]
\centering
\begin{tabular}{cccc}
\hline
\(d\) [mm] & \(\dot{m}\;\)[g\,s\(^{-1}\)] & \(v\;[\mathrm{cm}\,\mathrm{s}^{-1}]\) &
\(j\,\)[kg\,s\(^{-1}\,\)m\(^{-2}\)] \\ \hline
3 & 1.5 & 21.1 & 212 \\ 
3 & 2.0 & 28.1 & 283 \\
4 & 1.5 & 11.8 & 119 \\
4 & 2.0 & 15.8 & 159 \\ \hline
\end{tabular}
\caption{\label{tab:velocities}Flow speeds, \(v\), and mass velocities, \(j\),
  for the four different combinations of inner pipe diameter, \(d\), and
  mass-flow rate, \(\dot{m}\).}
\end{table}  
The flow speeds are computed using the relation
\[ v = \frac{4\,\dot{m}}{\pi\varrho_\mathrm{fl}\,d^2}\;. \]
The mass velocities are obtained according to \(j={\varrho_\mathrm{fl}}\,v\)
with \(\varrho_\mathrm{fl}=\SI{1007}{\kilogram\per\metre^3}\).

\begin{figure}
  \centering
  \includegraphics[width=\linewidth]{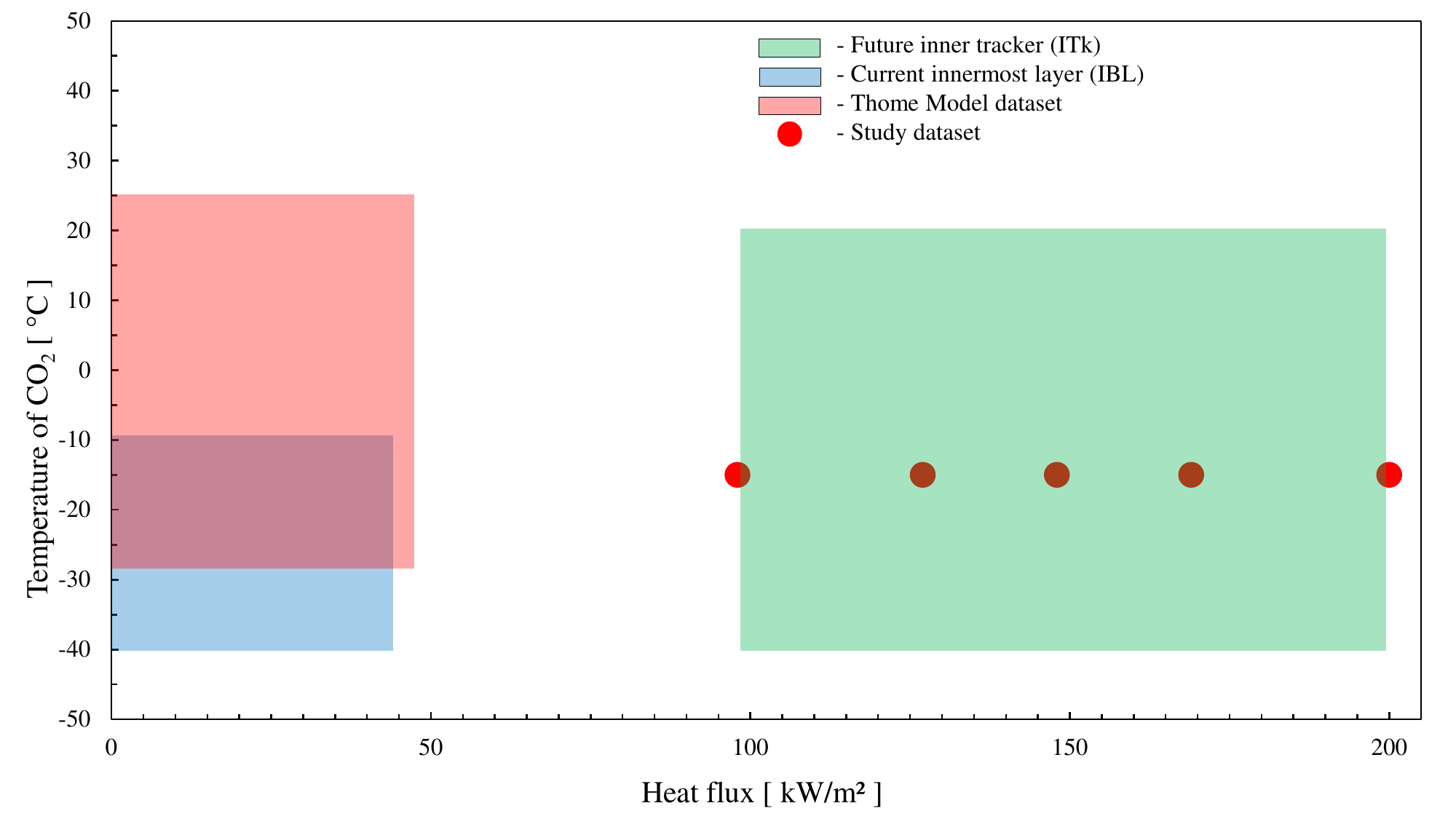}
  \caption{Operational range of the current and future cooling systems of the ATLAS
    pixel detector in the temperature-versus-heat-flux plane. In the presented study,
    the CO$_2$ was at an initial temperature of \(T=\SI{-15}{\degreeCelsius}\).
    Five different values of heat flux were applied, ranging from
    \SI{98.5}{kW\per\metre^2} to \SI{200}{kW\per\metre^2} and indicated by the red
    markers in the figure.
  }
  \label{fig:operation_range}
\end{figure}

Figure~\ref{fig:operation_range} shows the expected operating range of the CO$_2$
cooling of the future ATLAS Inner Tracker for the high luminosity LHC in the plane of
operation temperature versus heat flux (green area).
The heat flux ranges from below \SI{100}{kW\per\metre^2} to \SI{200}{kW\per\metre^2}
and the operation temperature from \SI{-35}{\degreeCelsius} to \SI{+30}{\degreeCelsius}.
For comparsion, the operating range of the innermost layer of the current
ATLAS Pixel detector, the Insertable B-layer~\cite{Barroca:2703341,ATLASIBL:2018gqd} (IBL), is
displayed as well (light blue area). 
The dataset of the Thome model~\cite{cheng2008} (red area) fits the operating
range of the IBL, but does not cover the range relevant for the new Inner Tracker.
This lack of data motivates the studies presented in this publication. 

In this study, the temperature was set to a fixed value of \(T=\SI{-15}{\degreeCelsius}\)
(\(=\SI{258.15}{\kelvin}\)).
To cover the range of heat flux predicted for the upgraded Inner Tracker
five different values were applied to the outer diameter of the pipes:
(98.5, 127, 149, 170, 200) kW/m$^2$.
However, for \(\dot{m}=\SI{1.5}{\gram\per\second}\) a heat flux of \SI{200}{kW\per\metre^2}
leads to dryout inside the pipe,
and thus only the four lower heat fluxes were applied in 
measurements with a mass-flow rate of \(\dot{m}=\SI{1.5}{\gram\per\second}\).
Lower mass-flow rates of \SI{0.5}{\gram\per\second} and \SI{1.0}{\gram\per\second}
were initially tested, but led to dryout of the pipe 
at even lower heat fluxes, and the measurements at these rates were thus discarded 
for further analysis. 

\begin{figure}
  \centering
  \includegraphics[width=\linewidth]{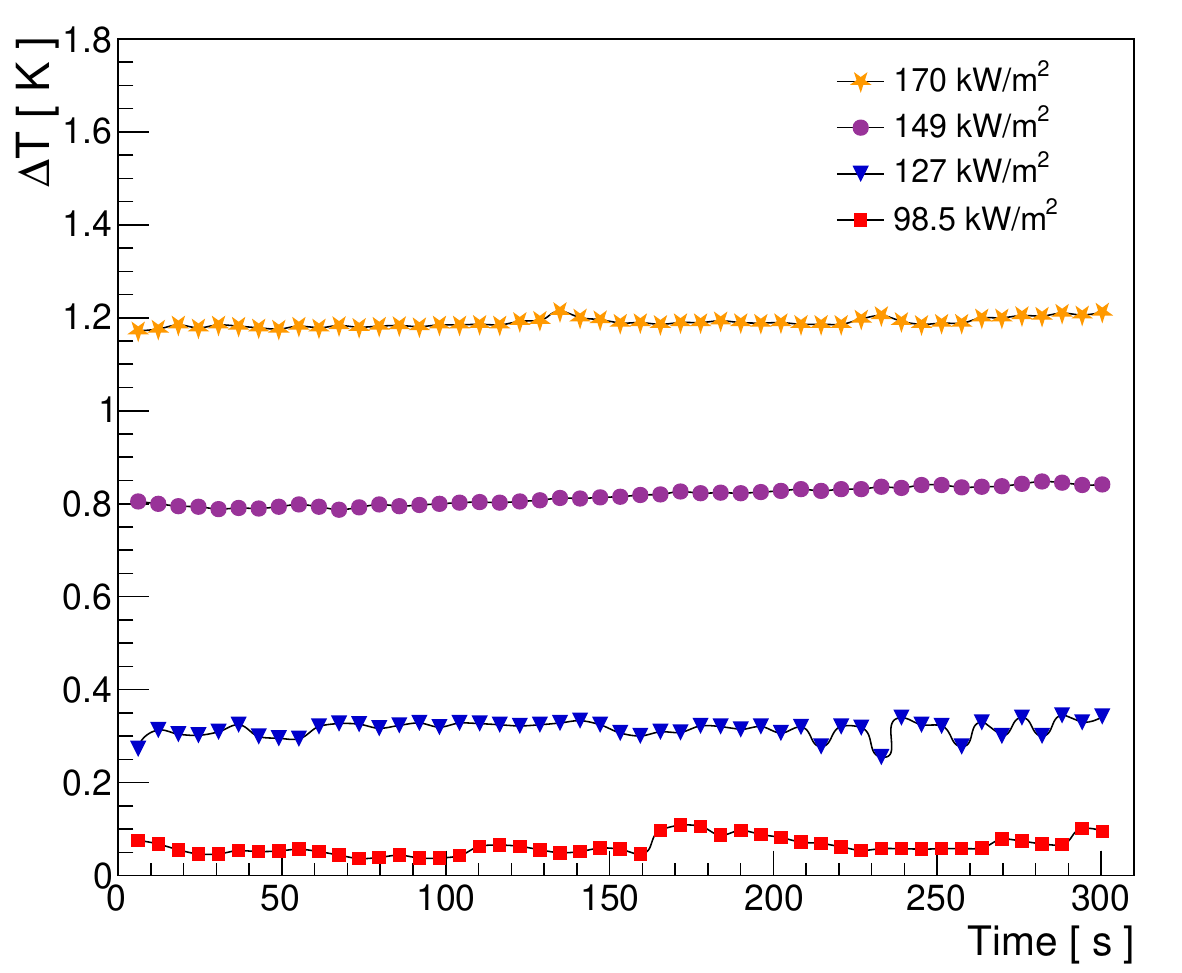}
  \caption{Example of the temperature increase \(\Delta T\) as a result of heating the
    pipe, obtained for measurements with a pipe of an inner diameter of \SI{3}{\milli\metre} mounted in
    horizontal orientation (\(\varphi=\ang{0}\)) and a mass-flow rate of
    \SI{1.5}{\gram\per\second}. Measurements with four different heat fluxes are displayed.
    The data points of the different measurement series are connected with a continuous line
    to guide the eye.}
  \label{fig:DeltaT}
\end{figure}

In total, 36 measurements with different parameters
(inner diameter, orientation, mass-flow rate, and heat flux) were performed.
Each measurement lasted 5 minutes. 
The temperatures in front of the heater and at the end of the prototype were recorded
once per second. 
As an example, Figure~\ref{fig:DeltaT} shows the temperature difference \(\Delta T\) before and
after heating the pipe as a function of time.
The measurement was done with a pipe of \(d=\SI{3}{\milli\metre}\), 
\(\varphi=\ang{0}\), and a mass-flow rate of \SI{1.5}{\gram\per\second}.
The \(\Delta T\) induced by the heating increases as expected for larger heat
flux and is nearly constant as a function of time, which indicates that
the measurement conditions were stable. 

\section{Determination of two-phase flow patterns and void fractions}
For different measurement setups different flow patterns are observed and identified.
The images recorded by the high-speed camera
are further used to determine the distribution of the void fraction. 
A pattern map is created based on these evaluations.

\subsection{Observation of flow patterns}
\begin{figure}[!t]
  \centering
  \subfloat[][]{
    \includegraphics[width=\linewidth]{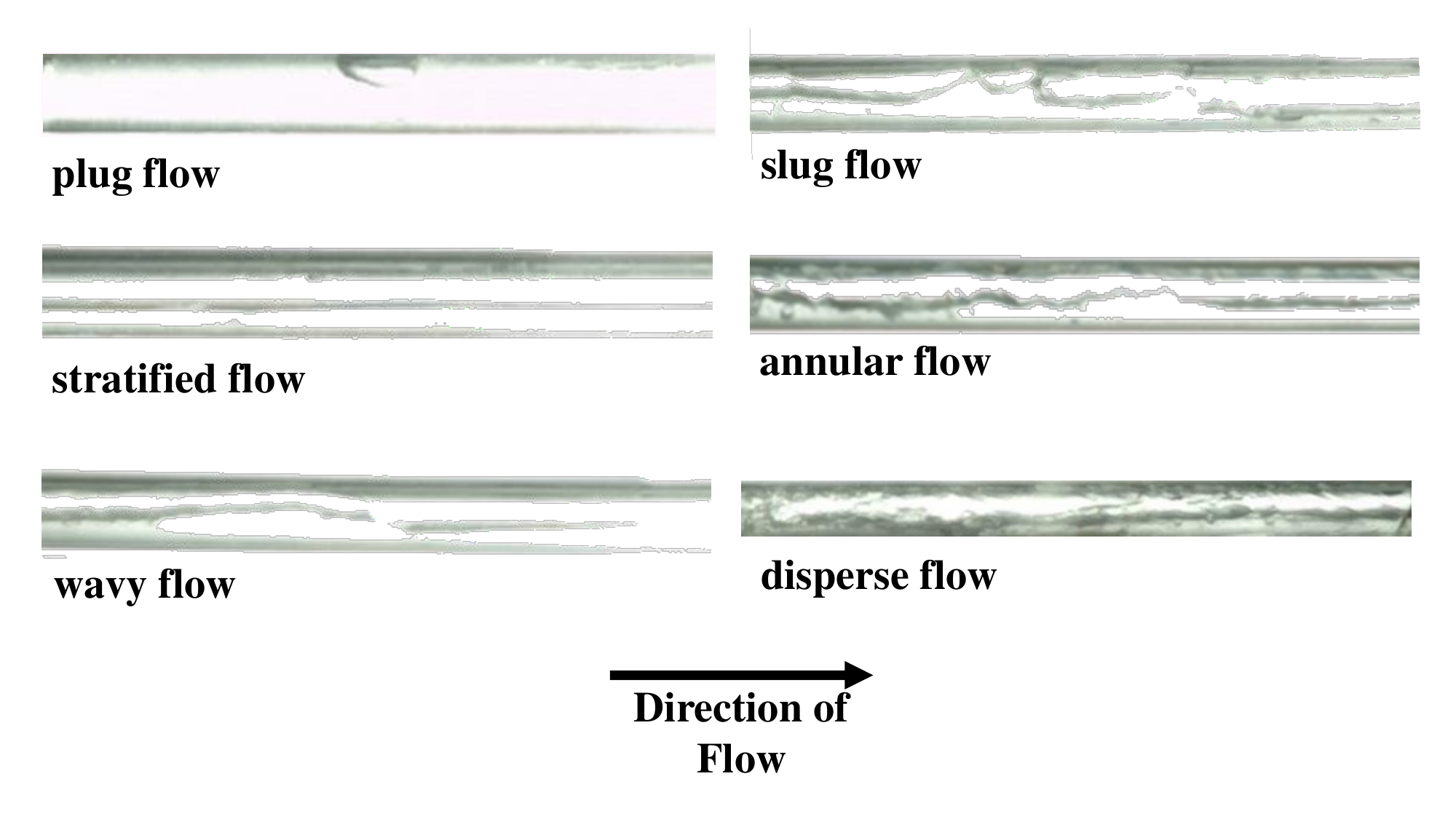}
    \label{subfig:flow_hori}
  }

  \subfloat[][]{
    \includegraphics[width=\linewidth]{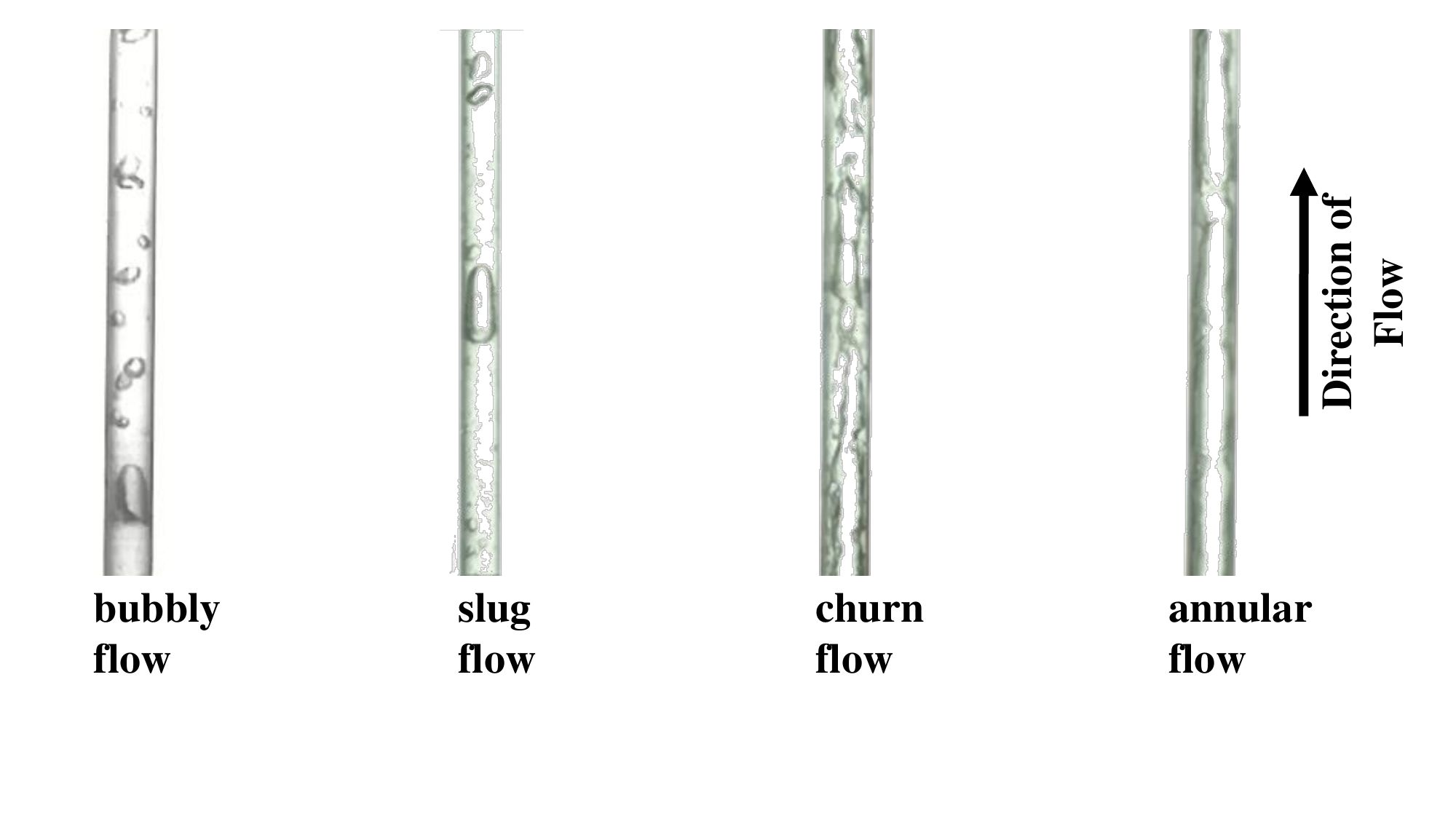}
    \label{subfig:flow_vert}
  }
  \caption{\label{fig:flowpattern}Flow patterns observed for \protect\subref{subfig:flow_hori}
    horizontal flow (\(\varphi=\ang{0}\)) and \protect\subref{subfig:flow_vert}
    vertical flow (\(\varphi=+\ang{90}\)).}
\end{figure}

Specific regularities in the distribution of gas and liquid phases of CO$_2$ inside the pipes 
are called flow patterns or flow regimes.
Flow patterns are identified by visual evaluation following the classification in Ref.~\cite{Brennen:2005}.
In this experimental study, flow patterns were observed and identified as shown in
Fig.~\ref{fig:flowpattern} for both horizontal and vertical flow.
The classification scheme comprises the following categories:
\paragraph{Bubbly flow} Inside the pipe, the gas phase is visible as bubbles.
Although the size of the bubbles varies, the overall shape is spherical.
The liquid phase of CO$_2$ surrounds the gaseous phase.
This pattern was observed for vertical flow, see Fig.~\ref{fig:flowpattern}\subref{subfig:flow_vert},
but not for horizontal flow.
\paragraph{Plug flow} As shown in Fig.~\ref{fig:flowpattern}\subref{subfig:flow_hori},
plug flow features gas fragments that are shaped like a long bubble and are separated from the
liquid phase inside the pipe. The size of the plugs is smaller than the inner pipe diameter. 
\paragraph{Stratified flow} Liquid and gas are completely divided inside the pipe. The gas phase
is in the horizontal orientation on top and the liquid phase is at the bottom of the pipe.
This pattern is only relevant for the horizontal flow and was observed in many setups 
in the studies with \(\varphi=\ang{0}\),
see Fig.~\ref{fig:flowpattern}\subref{subfig:flow_hori}.
\paragraph{Wavy flow} The stratified flow creates extra waves between the gas and liquid phases
as the velocity rises. The waves never reach the top of the pipe.
\paragraph{Slug flow} The produced gas bubbles have nearly the same size as the inner diameter of
the pipe. The liquid phase of CO$_2$ divides the gaseous phase, separating the long-stretched
gas bubbles. The slug flow pattern was observed for the horizontal (\(\varphi=\ang{0}\))
as well as the vertical (\(\varphi=+\ang{90}\)) orientation of the pipe.
\paragraph{Churn flow} The slugs of the gaseous phase are dissolving and become an almost
continuous gas flow. Churn flow is a highly disturbed flow of two phases.
Increasing the velocity of a flow causes the structure to become unstable. Churn flow is
characterized by the presence of a very thick and unstable fluid film,
with the fluid oscillating up and down frequently.
\paragraph{Annular flow} The CO$_2$ forms a constant annular liquid-phase film along the pipe wall.
The liquid phase is thicker at the bottom than at the top. The edge between the gas phase and
the liquid phase is mixed up with small waves and little drops. 
\paragraph{Disperse flow} At high gas-flow rates, the liquid-phase film becomes thinner until
it decreases to single droplets in a constant gas phase. The pipe surface is in direct contact
with the gas phase and the previously existing annular film has dried out~\cite{Brennen:2005}. 

\vspace*{10mm}

As indicated in Fig.~\ref{fig:flowpattern}, the two-phase flow patterns described above were
observed in the studies performed for different measurement parameters. Table~\ref{tab:flow_matrices}
provides an overview on which flow pattern occured for different flow speeds and heat fluxes
for horizontal and vertical flow.
\begin{table*}[t]
\centering
  \begin{tabular}{cccccc}
  \hline
  \multicolumn{6}{l}{Horizontal flow (\(\varphi=\ang{0}\))}  \\
   & \multicolumn{5}{c}{Heat flux [ kW/m\(^{2}\) ]} \\
  \(v\;[\mathrm{cm}\,\mathrm{s}^{-1}]\) & 98.5 & 127 & 149 & 170 & 200 \\
  \hline
  11.8 & stratified--wavy & stratified--wavy & stratified--wavy & slug  & disperse \\
  15.8 & plug--stratified & stratified--wavy & stratified--wavy & slug & slug \\
  21.1 & wavy--slug       & slug       & slug       & slug & dryout \\
  28.1 & stratified & stratified--wavy & stratified--wavy & slug & disperse \\
  \hline
  & & & & & \\ \hline
  \multicolumn{6}{l}{Vertical flow (\(\varphi=+\ang{90}\))} \\
  & \multicolumn{5}{c}{Heat flux [ kW/m\(^{2}\) ]} \\
  \(v\;[\mathrm{cm}\,\mathrm{s}^{-1}]\) & 98.5 & 127 & 149 & 170 & 200 \\ \hline
  11.8 & slug & slug & churn & churn & dryout \\
  15.8 & slug & slug & churn & churn & churn \\
  21.1 & slug & churn & churn & churn & dryout \\
  28.1 & slug & churn & churn & annular & annular \\ \hline
  \end{tabular}
\caption{\label{tab:flow_matrices}Observed flow patterns for different flow speeds \(v\) and
  heat fluxes for horizontal flow (\(\varphi=\ang{0}\))
  and vertical flow (\(\varphi=+\ang{90}\)).}
\end{table*}
In the case of horizontal flow, stratified flow occurs predominantly for heat fluxes from
\SI{98.5}{kW\per\metre^2} to \SI{149}{kW\per\metre^2}, while slug flow is the dominating
pattern for higher heat fluxes.
In vertical pipes, slug flow dominates at low heat fluxes and churn flow is the typical
pattern for higher heat fluxes.
  
\subsection{Determination of the void fraction}
\begin{figure}[!tbh] 
  \centering
    \includegraphics[width=0.95\linewidth]{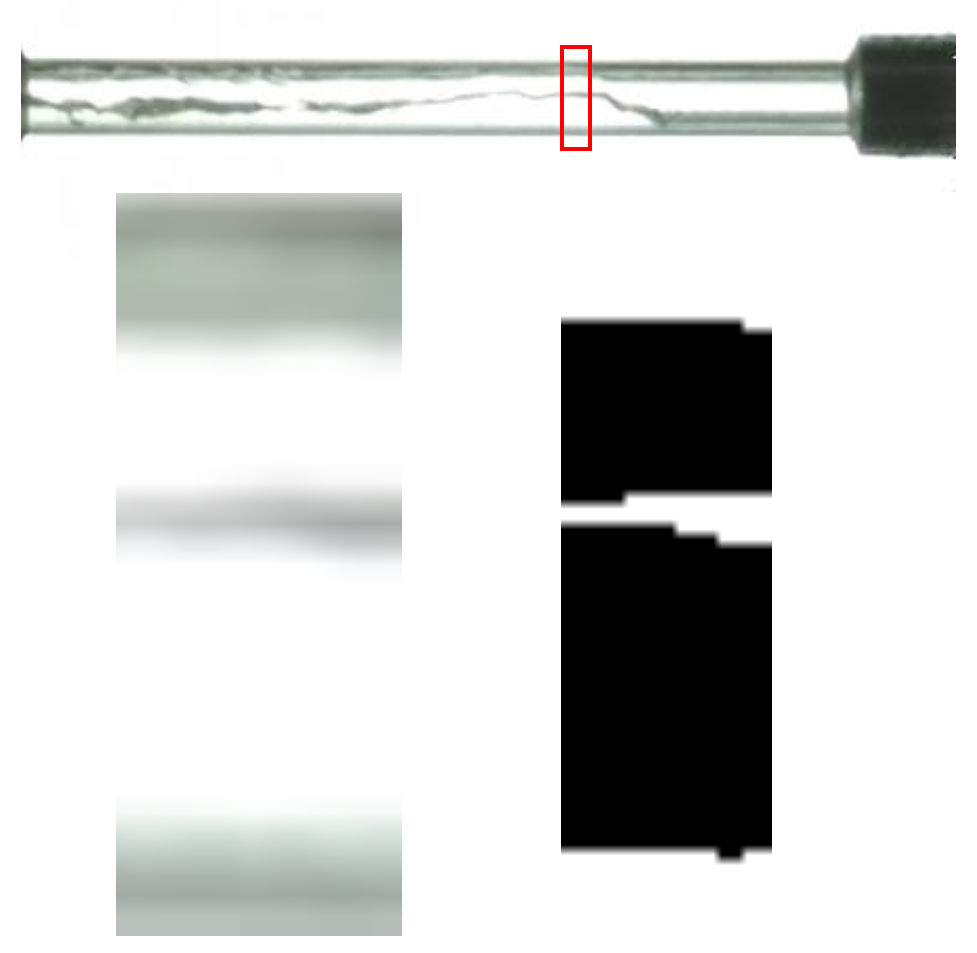}
  \caption{\label{fig:imagine_conversion}Visualisation of the conversion of the
    recorded image into a binary image in order to determine the void fraction.
    On the top the complete recorded picture is shown. 
    On the lower left ten neighbouring pixel columns are shown
    for better illustration of the method, even though the analysis uses only one
    of the columns. The method relies on determining the interface between the
    gaseous and the liquid phase. In the recorded image the interface is
    characterised by a darker colour because of reduced light transition due to
    scattering. In the binary image the transition region is given by the white
    strip between the two black areas. 
  }
\end{figure}

A Python application employing image-processing tools was created to analyse the recorded images
and determine the void fraction. The size of the gaseous phase is determined in a one-pixel-broad
column of the recorded image at one specific location along the vision block for each measurement.
The measurement location is defined to be in the last third of the block, in order to increase
the distance to the heater, and
is set for each measurement individually, using a point of clear vision.
The image-transform method {\it Otsu} is employed to convert the cropped column of the image into a binary
image, as illustrated in Figure~\ref{fig:imagine_conversion}.
Subsequently, the edge-detection method is used to identify the interface 
separating the gas and the liquid phase.
For horizontal flow only one interface is assumed because the gas phase occurs
always on top of the pipe. When analysing images of vertical flow, two
interfaces are considered. 
Dividing the determined size of the gaseous phase by the inner diameter of the pipe gives the void fraction.

\begin{figure*}[!h]
  \centering
  \includegraphics[width=0.62\linewidth]{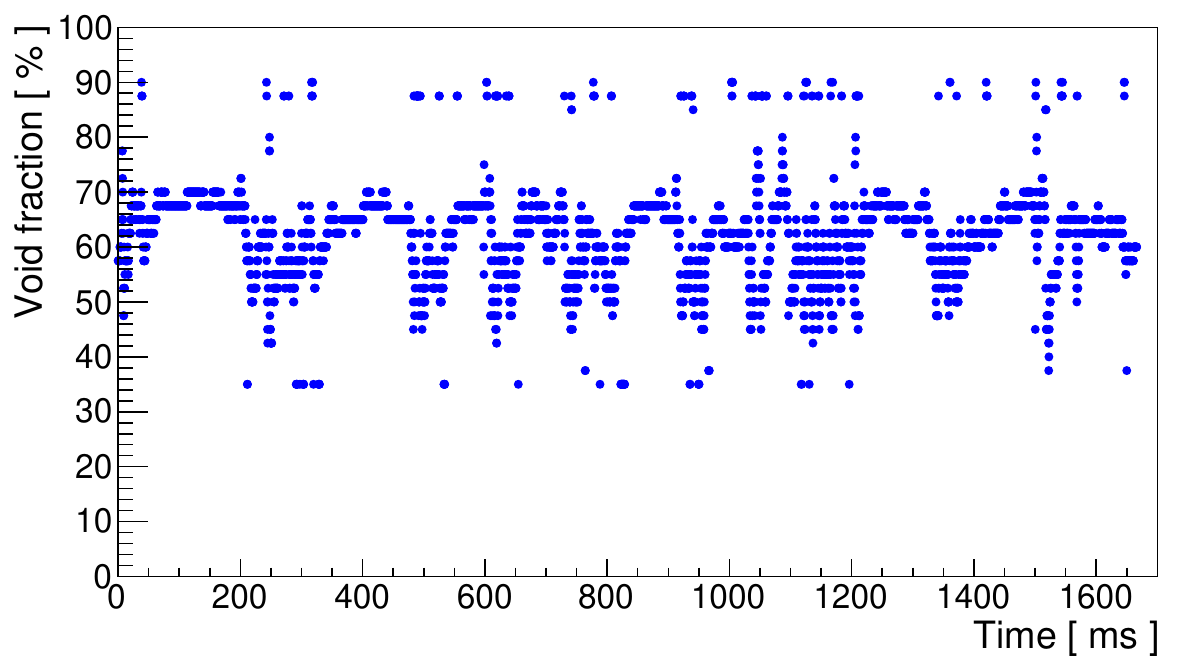}
  \caption{Void fraction at a fixed position of the vision block as a function of time
    for a measurement with a pipe of an inner diameter of \SI{3}{\milli\metre} mounted in
    horizontal orientation (\(\varphi=\ang{0}\)), a mass-flow rate of
    \SI{1.5}{\gram\per\second} and a heat flux of \SI{98.5}{kW\per\metre^2}.}
  \label{fig:gas_fraction}
\end{figure*}
The automised method of determining the void fraction was validated by manually analysing
randomly chosen images.
Both methods were found to agree in determining the void fraction within \(\pm5\%\).
This value is taken as a systematic uncertainty in determining the void fraction.  

The code consecutively scans all \num{5000} images of a single measurement and
determines a value of the void fraction for each image, leading to an ntuple of
\num{5000} void-fraction values for each measurement.

As an example, Figure~\ref{fig:gas_fraction} shows the change of the void fraction at a fixed position
of the vision block as a function of time for a measurement with a pipe of an inner diameter
of \SI{3}{\milli\metre} mounted in horizontal orientation (\(\varphi=\ang{0}\)). A mass-flow rate of
\SI{1.5}{\gram\per\second} and a heat flux of \SI{98.5}{kW\per\metre^2} were used for this
measurement.
The void fraction varies in a range between approximately 45\% and 70\%,
with a few fluctuations to 35\% on the low side and 90\% on the high end. 

\subsection{Distributions of void fractions}
\label{sec:dist_gas_frac}
To compare the void fractions observed in different measurements with each other,
distributions of void fractions are formed by filling the measured void fractions into histograms. 
A bin width of 10\% is chosen for the histograms in order to account for the systematic uncertainty of the 
determined void fractions.
\begin{figure*}[!thb]
  \centering
  \subfloat[][$d=\SI{3}{\milli\metre}$ and $\dot{m}=\SI{1.5}{\gram\per\second}$]{
  \includegraphics[width=0.47\linewidth]{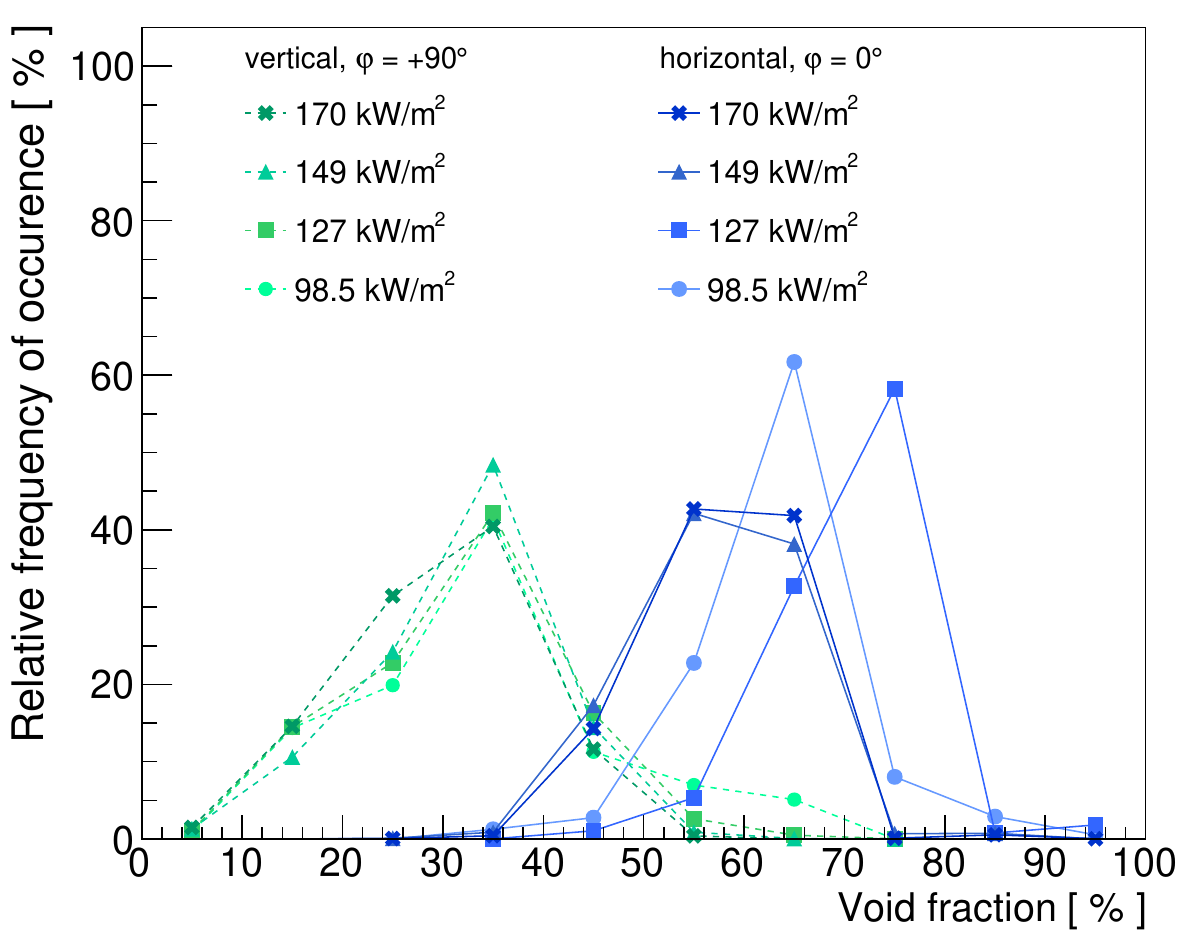}
    \label{subfig:dist_3mm_1_5gs}  
  }
  \quad
  \subfloat[][$d=\SI{4}{\milli\metre}$ and $\dot{m}=\SI{1.5}{\gram\per\second}$]{
  \includegraphics[width=0.47\linewidth]{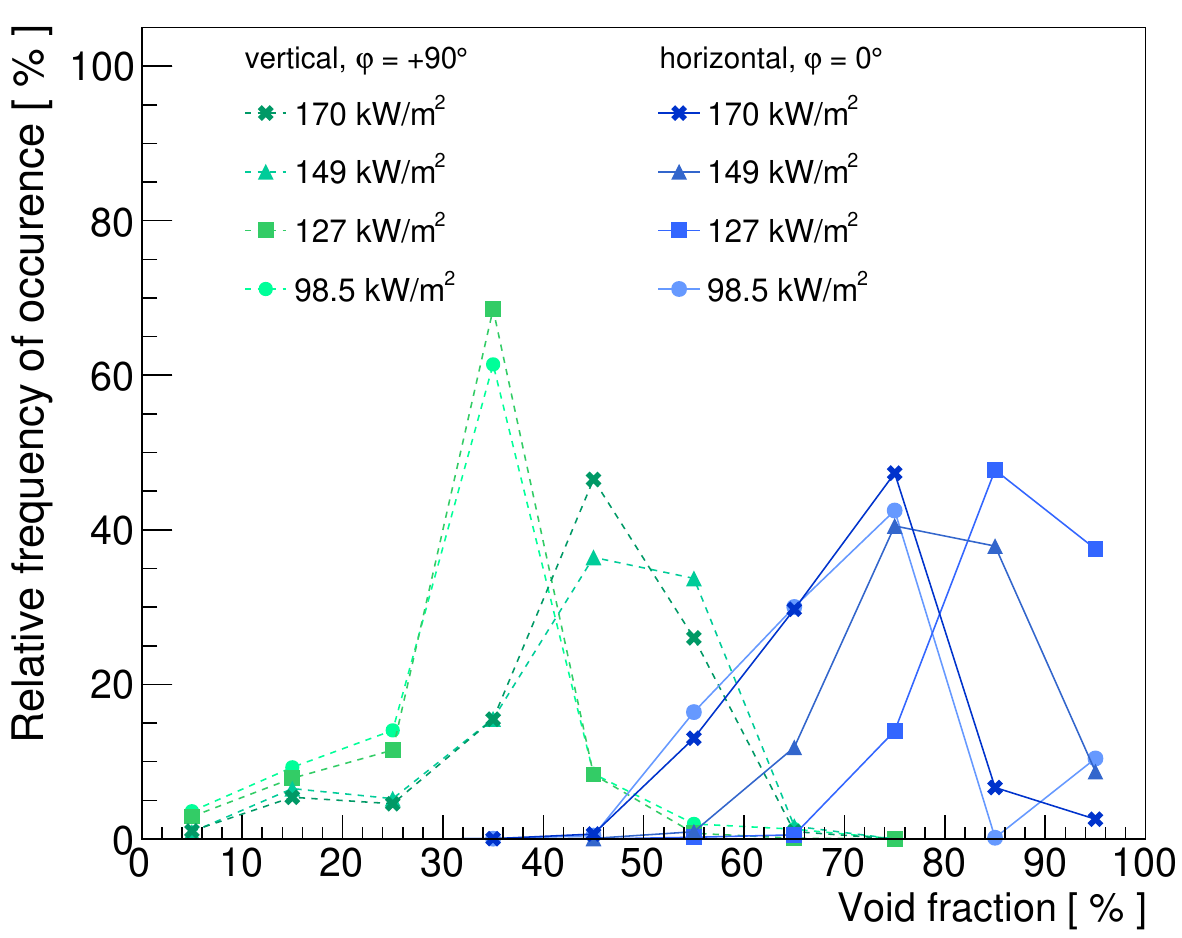}
    \label{subfig:dist_4mm_1_5gs}
  }  

  \subfloat[][$d=\SI{3}{\milli\metre}$ and $\dot{m}=\SI{2}{\gram\per\second}$]{
  \includegraphics[width=0.47\linewidth]{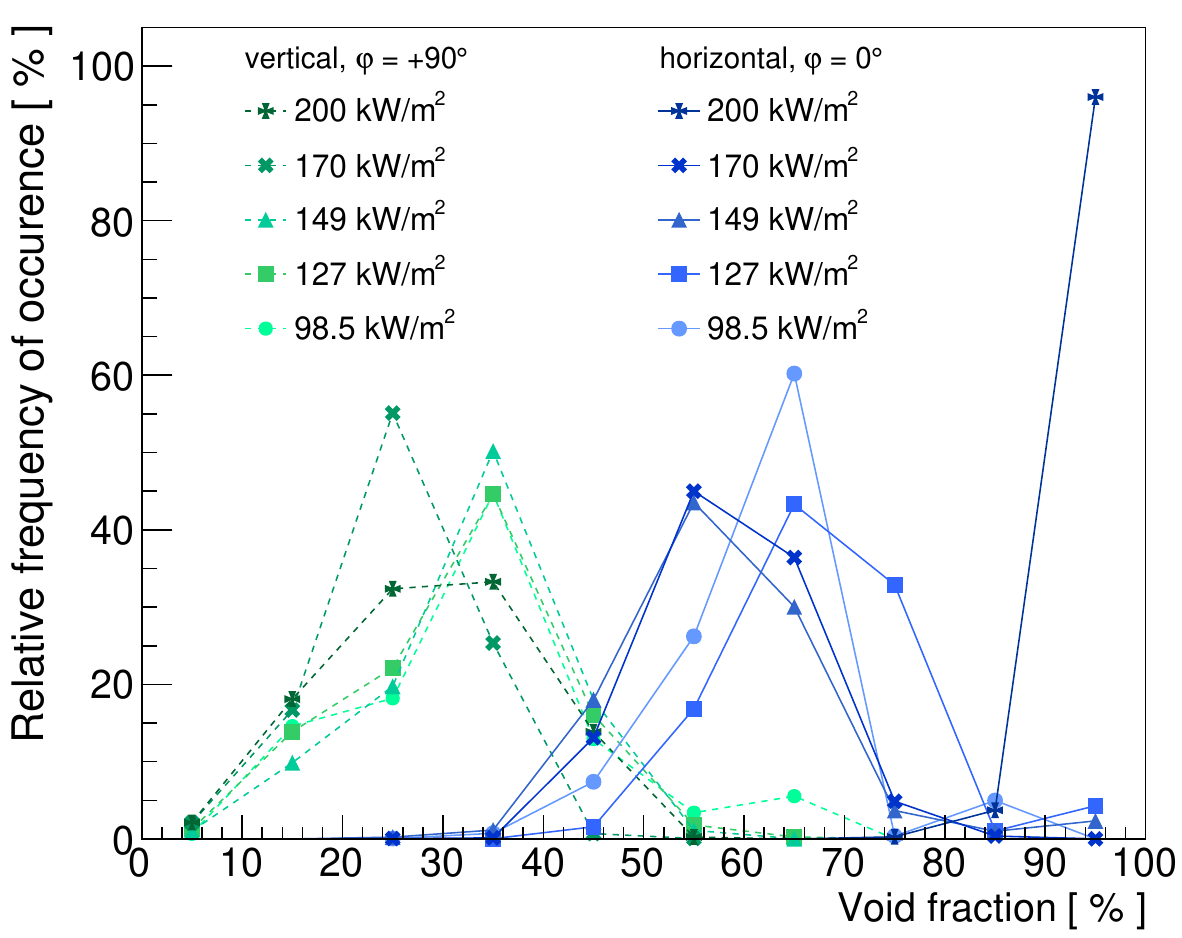}
    \label{subfig:dist_3mm_2gs}
  }
  \quad
  \subfloat[][$d=\SI{4}{\milli\metre}$ and $\dot{m}=\SI{2}{\gram\per\second}$]{
  \includegraphics[width=0.47\linewidth]{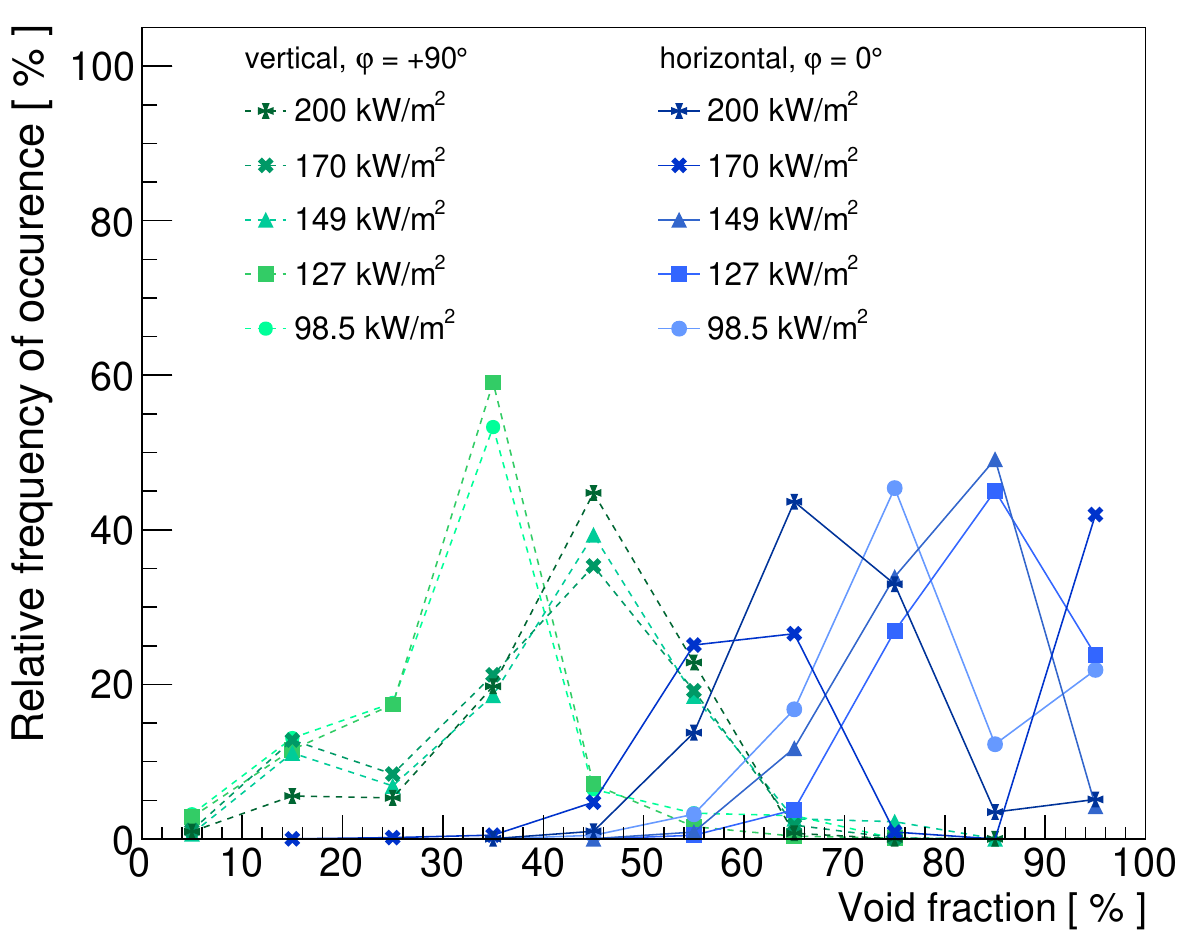}
    \label{subfig:dist_4mm_2gs}
  }
  \caption{Distribution of void fractions for measurements with a pipe of an inner
    diameter of \SI{3}{\milli\metre} in \protect\subref{subfig:dist_3mm_1_5gs} and
    \protect\subref{subfig:dist_3mm_2gs}, and \SI{4}{\milli\metre} in
    \protect\subref{subfig:dist_4mm_1_5gs} and \protect\subref{subfig:dist_4mm_2gs}.
    Mass-flow rates of $\dot{m}=\SI{1.5}{\gram\per\second}$, \protect \subref{subfig:dist_3mm_1_5gs} and
    \protect\subref{subfig:dist_4mm_1_5gs}, and $\dot{m}=\SI{2}{\gram\per\second}$,
    \protect\subref{subfig:dist_3mm_2gs} and \protect\subref{subfig:dist_4mm_2gs}, are applied. 
    The results are given for a series of measurements
    with a horizontal orientation of the pipe (\(\varphi=\ang{0}\)), displayed with blue markers, 
    and measurements with a vertical orientation of the pipe and upward flow
    (\(\varphi=+\ang{90}\)), displayed with green markers.
  }
  \label{fig:dist_gas_frac}
\end{figure*}
Figure~\ref{fig:dist_gas_frac} shows the distributions of void fractions for all
36 measurement setups under investigation.
Measurements of vertical flow are represented by green markers connected with dashed lines,
measurements of horizontal flow are displayed with blue markers connected by full lines.
The statistical uncertainties of the data points are equal or below the marker size and are therefore
not shown.

Clear differences in void fraction are observed between vertical and horizontal flow
for all values of heat flux.
The distributions of the void fraction tend to larger values for horizontal flow than for vertical flow 
and show larger variability for different heat fluxes.
For vertical flow the maxima of the distributions are between 30\% and 50\%,
for horizontal flow between 50\% and 100\%. 
For a pipe with an inner diameter of \SI{3}{\milli\metre} and a mass-flow rate
of $\dot{m}=\SI{1.5}{\gram\per\second}$, see Fig.~\ref{fig:dist_gas_frac}\subref{subfig:dist_3mm_1_5gs},
the maxima of the distributions of vertical flow are located between 30\% and 40\%,
while the maxima of the measurements of horizontal flow are between 50\% and 80\%,
spreading over a larger range.
For the measurements with pipes of an inner diameter of \SI{4}{\milli\metre} and 
\(\varphi=+\ang{90}\),
see Figs.~\ref{fig:dist_gas_frac}\subref{subfig:dist_4mm_1_5gs} and
\ref{fig:dist_gas_frac}\subref{subfig:dist_4mm_2gs}, the distributions peak between 
30\% and 40\% for heat fluxes of \SI{98.5}{kW\per\metre^2} and
\SI{127}{kW\per\metre^2}, while the distributions for higher heat fluxes have a
maximum between 40\% and 50\%. However, this trend to higher void fractions for higher heat
fluxes is not seen in Figure~\ref{fig:dist_gas_frac}\subref{subfig:dist_3mm_2gs}.

\begin{figure}[t]
  \centering
  \includegraphics[width=0.88\linewidth]{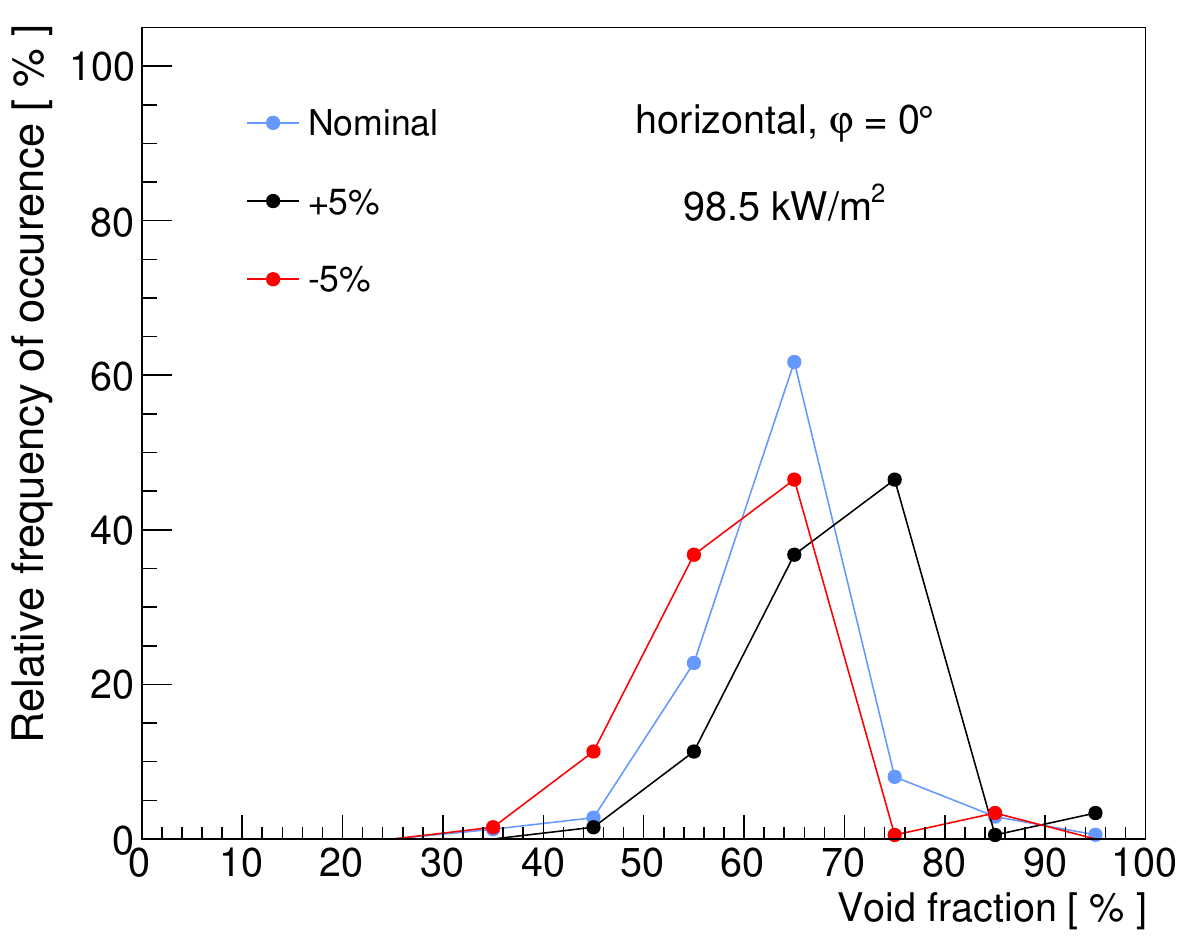}
  \caption{\label{fig:syst_impact}Illustration of the impact of a systematic uncertainty in
    the form of a coherent absolute shift of \(\pm5\%\) to the measured void fraction 
    on the corresponding distribution observed for horizontal flow at a heat flux of
    \SI{98.5}{kW\per\metre^2} in a pipe with an inner diameter of \SI{3}{\milli\metre} and
    a mass-flow rate of $\dot{m}=\SI{1.5}{\gram\per\second}$.}
\end{figure}

The systematic uncertainty in determining the void fraction, estimated to be \(\pm5\%\),
induces a systematic shift of the distributions along the \(x\)-axis if a coherent shift
of void-fraction measurements is assumed.
An illustration of this effect is given in Fig.~\ref{fig:syst_impact}.
The black markers represent the void-fraction distribution for a coherent shift of \(+5\%\), 
while the red markers indicate the distribution obtained for a shift of the measured 
values by \(-5\%\). 
Based on this consideration, the distributions observed for different heat fluxes in a
specific orientation of the pipe can be seen as compatible within the uncertainties.
However, the distributions of void fractions for vertical and horizontal orientations 
of the pipe show a genuine difference, even if a systematic uncertainty of \(\pm5\%\) is
taken into account. This finding underlines the importance of performing separate
measurements in both orientations.

The differences between horizontal and vertical flow are due to the effect of gravity
which causes the gas phase to gather in the upper part of the pipe in the case of
horizontal flow, favouring the clear division of gaseous and fluid components, and
thus leading to the formation of clearly distinct gas bubbles,
while for vertical flow gas may remain scattered across the pipe in tiny bubbles
which are not detectable with the given optical resolution.
Therefore, the pipe wall is covered with fluid to a large extent in the case of vertical flow,
leading to larger friction compared to horizontal flow. 
In addition, in vertical orientation the gas bubbles are subject to a
buoyance force which may cause them to travel at a faster upward speed than the fluid,
and may thus lead to a reduction of the void fraction.
\FloatBarrier

\subsection{Two-phase flow-pattern map}
\begin{figure}[t]
  \centering
  \subfloat[][]{
    \includegraphics[width=0.85\linewidth]{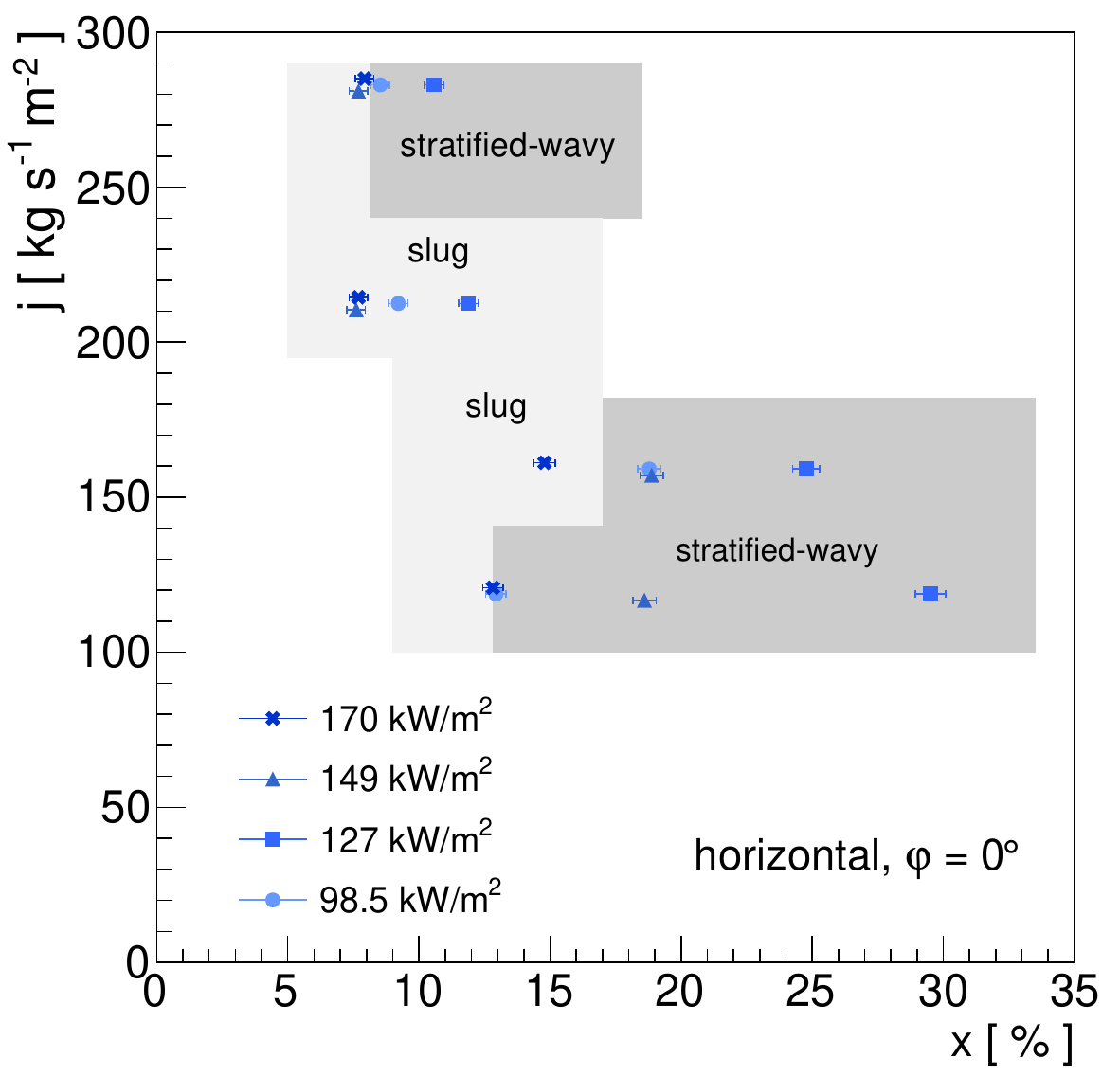}
    \label{subfig:pattern_map_horizontal}
  }
  
  \subfloat[][]{
    \includegraphics[width=0.85\linewidth]{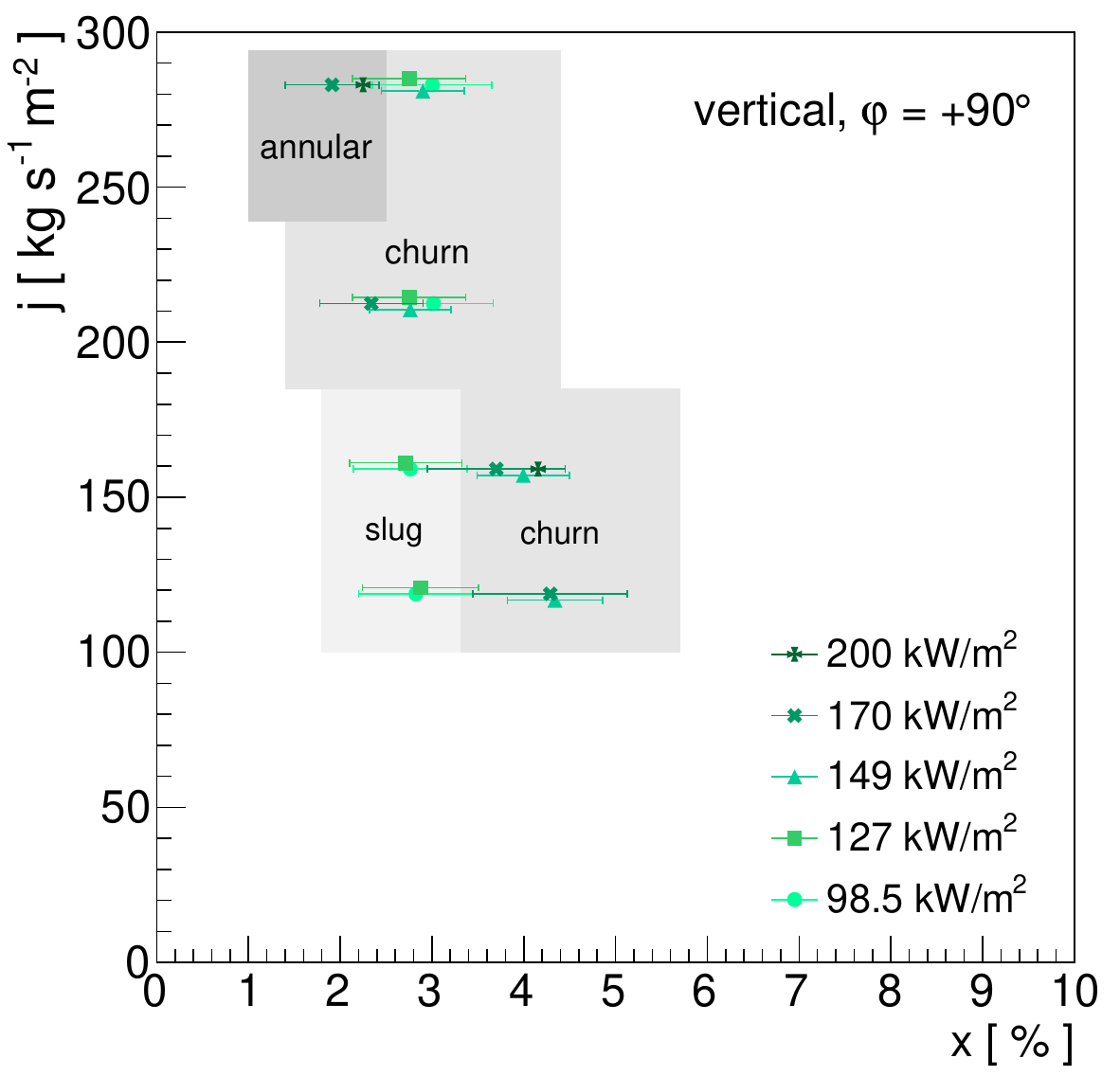}
    \label{subfig:pattern_map_vertical}
  }
  \caption{\label{fig:pattern_maps}Pattern maps of two-phase flow of CO$_2$. Flow patterns are
    identified in the plane of mass velocity \(j\) versus the vapour quality \(x\).
    Different flow speeds are realised by different mass-flow rates and different
    inner diameters of the pipe as listed in Table~\ref{tab:velocities}.
    The map of patterns observed in horizontal flow with \(\varphi=\ang{0}\) are shown in
    \protect\subref{subfig:pattern_map_horizontal} and
    the patterns found in vertical flow (\(\varphi=+\ang{90}\)) in \protect\subref{subfig:pattern_map_vertical}.
    To better display the different data points and their error bars, small offsets are added or
    subtracted from the values of the flow speeds.
    Measurements with different heat flux and orientation are shown.
    For horizontal flow the measurements with a heat flux of \SI{200}{kW\per\metre^2} are not
    included because for three out of four measurements disperse flow or dryout were found.
  }
\end{figure}

Following Ref.~\cite{cheng2008}, two-phase flow-pattern maps are created based on the
determination of the average void fraction, named \(\bar{\alpha}\),
and the evaluation of the recorded images concerning the flow patterns. 
The data (ntuples) presented in Fig.~\ref{fig:dist_gas_frac} as histograms are used to determine
\(\bar{\alpha}\). The assignment of flow patterns is taken from
Table~\ref{tab:flow_matrices}.
The determined values of \(\bar{\alpha}\) are turned into the vapour quality \(x\) based on
the relation
\[x=\frac{\varrho_\mathrm{g}\cdot\bar{\alpha}}
         {\varrho_\mathrm{g}\cdot\bar{\alpha}+\varrho_\mathrm{fl}\cdot(1-\bar{\alpha})}, \]
using \(\varrho_\mathrm{g}=\SI{60.8}{\kilogram\per\metre^3}\).
The maps present the measurements as markers in the plane of mass velocity \(j\)
versus vapour quality \(x\) and are shown in Figure~\ref{fig:pattern_maps},
separately for horizontal flow (\(\varphi=\ang{0}\)) and vertical flow (\(\varphi=+\ang{90}\)).

As already observed in Section~\ref{sec:dist_gas_frac}, the two sets of measurements related to 
horizontal flow and vertical flow form distinct populations in the pattern maps, and are thus
displayed in separate diagrams. For horizontal flow vapour qualities between 7\% and 30\% are found,
while for vertical flow values between 2\% and 4\% are observed.
For horizontal flow the gravitational force favours the clear 
division of gaseous and fluid components and leads to the formation of distinct large gas bubbles.
The flow speed influences the vapour quality as well: setups with higher mass veloctiy tend to have
smaller vapour quality. This dependence is visible for both vertical and horizontal flow.

\section{Measurements of pressure drop}
Frictional forces lead to a pressure drop when a fluid moves a along a pipe.  
The formation of a gaseous phase in fluid CO$_2$ due to the absorption of heat results in an
additional dynmic pressure drop. In vertical pipes, the static (gravitational) pressure drop is added. 
\begin{figure*}[!tbh]
  \centering
  \subfloat[][]{
    \includegraphics[width=0.45\linewidth]{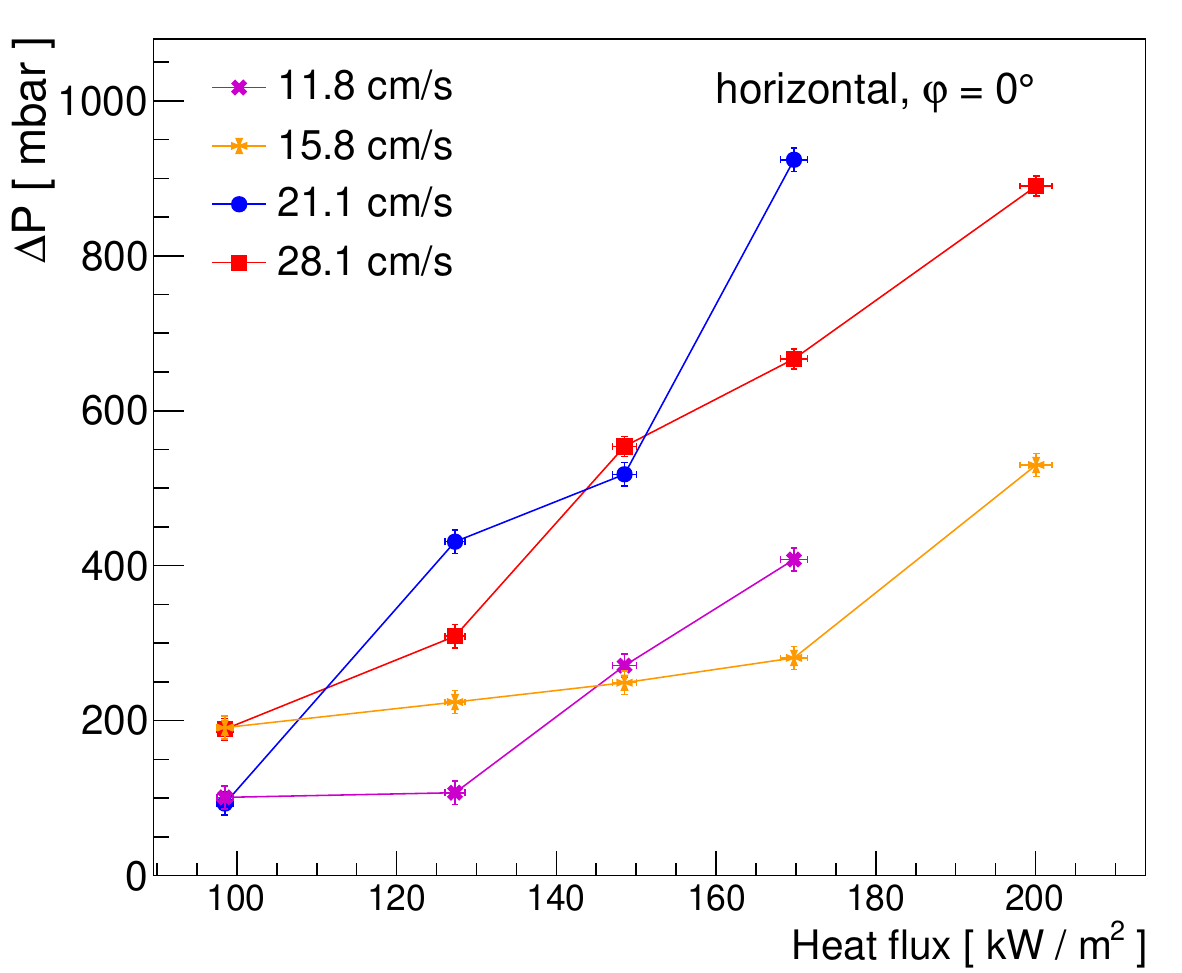}
    \label{subfig:deltaP_heatflux_phi00}
  }
  \quad
  \subfloat[][]{
    \includegraphics[width=0.45\linewidth]{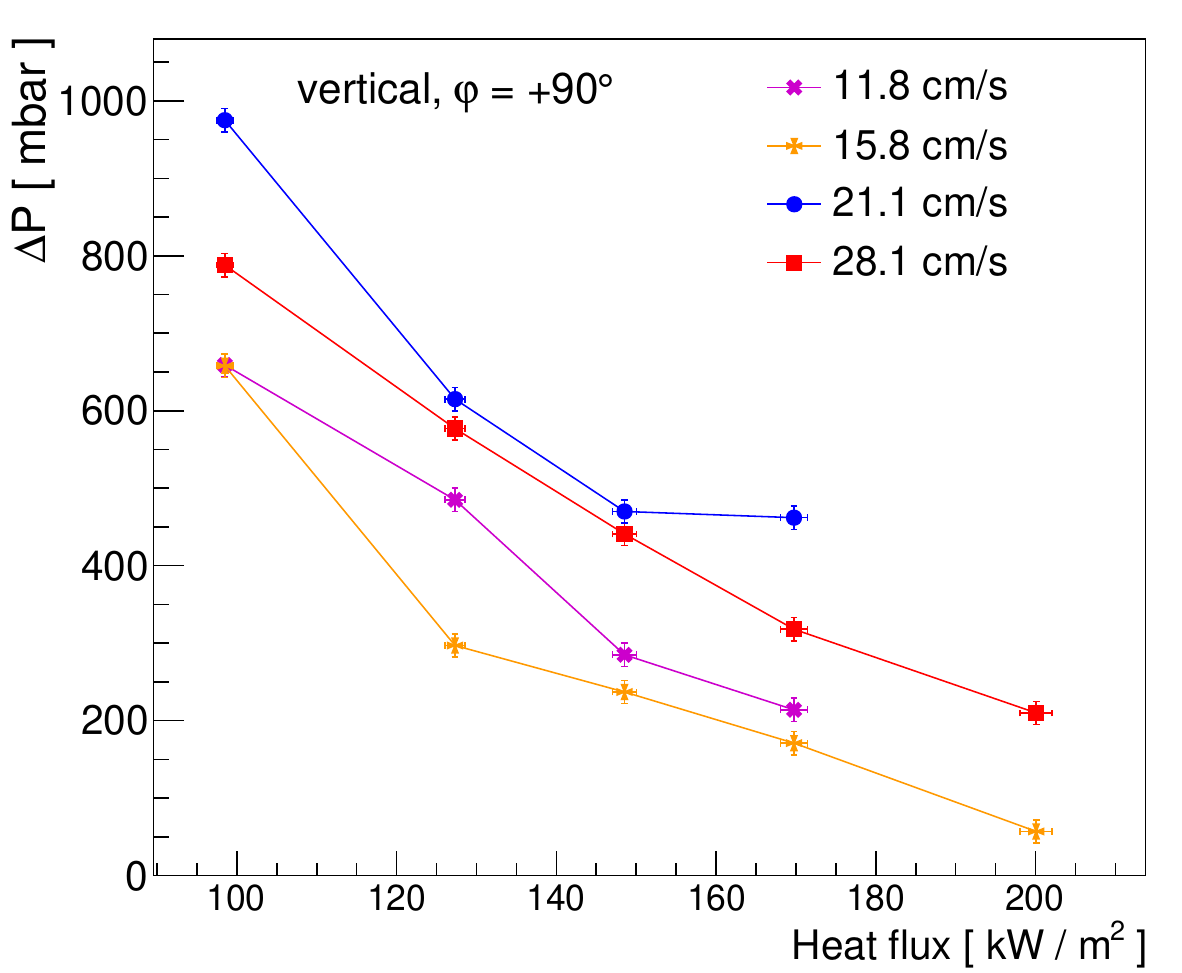}
    \label{subfig:deltaP_heatflux_phi90}
  }

  \subfloat[][]{
     \includegraphics[width=0.45\linewidth]{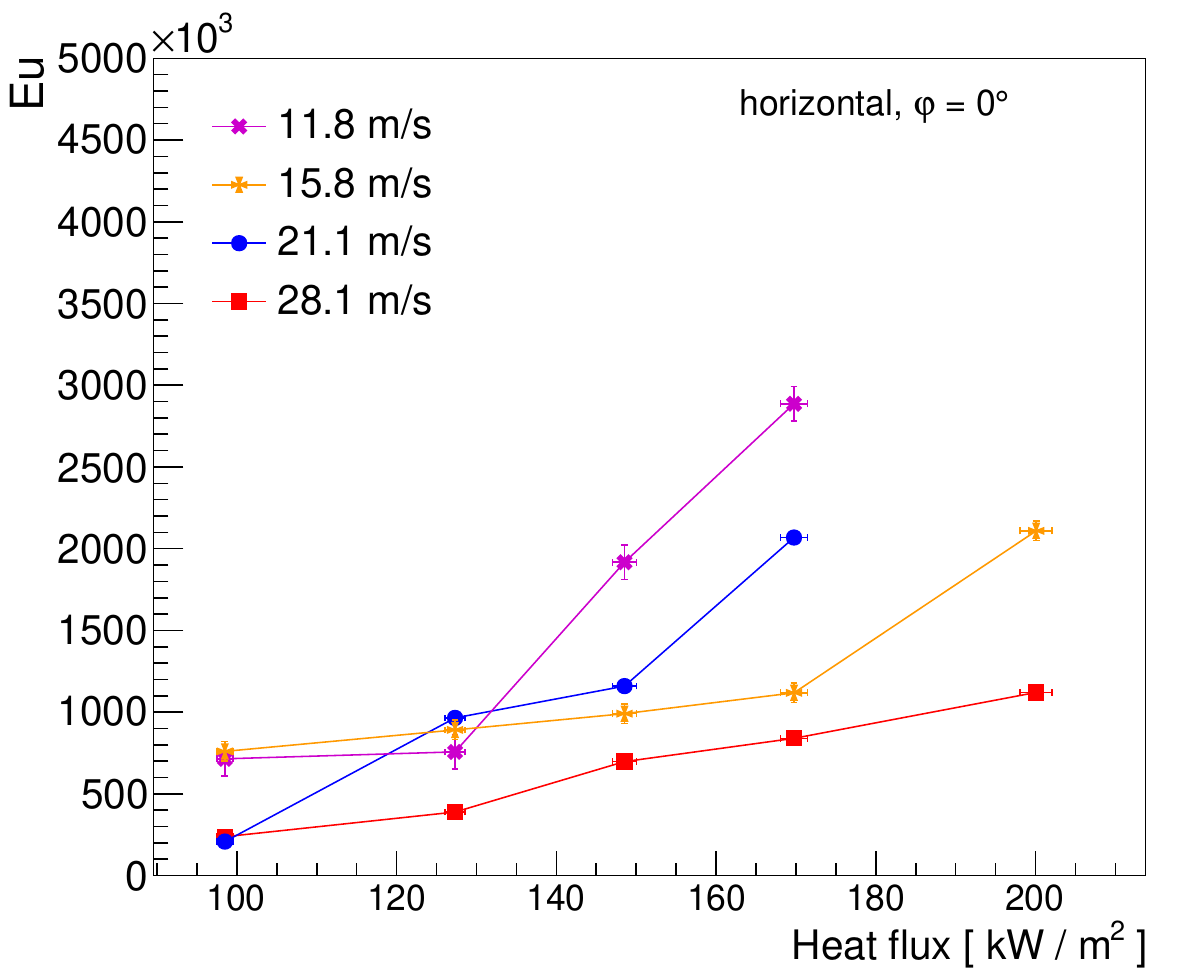}
     \label{subfig:Eu_heatflux_phi00}
   }
  \quad
  \subfloat[][]{
     \includegraphics[width=0.45\linewidth]{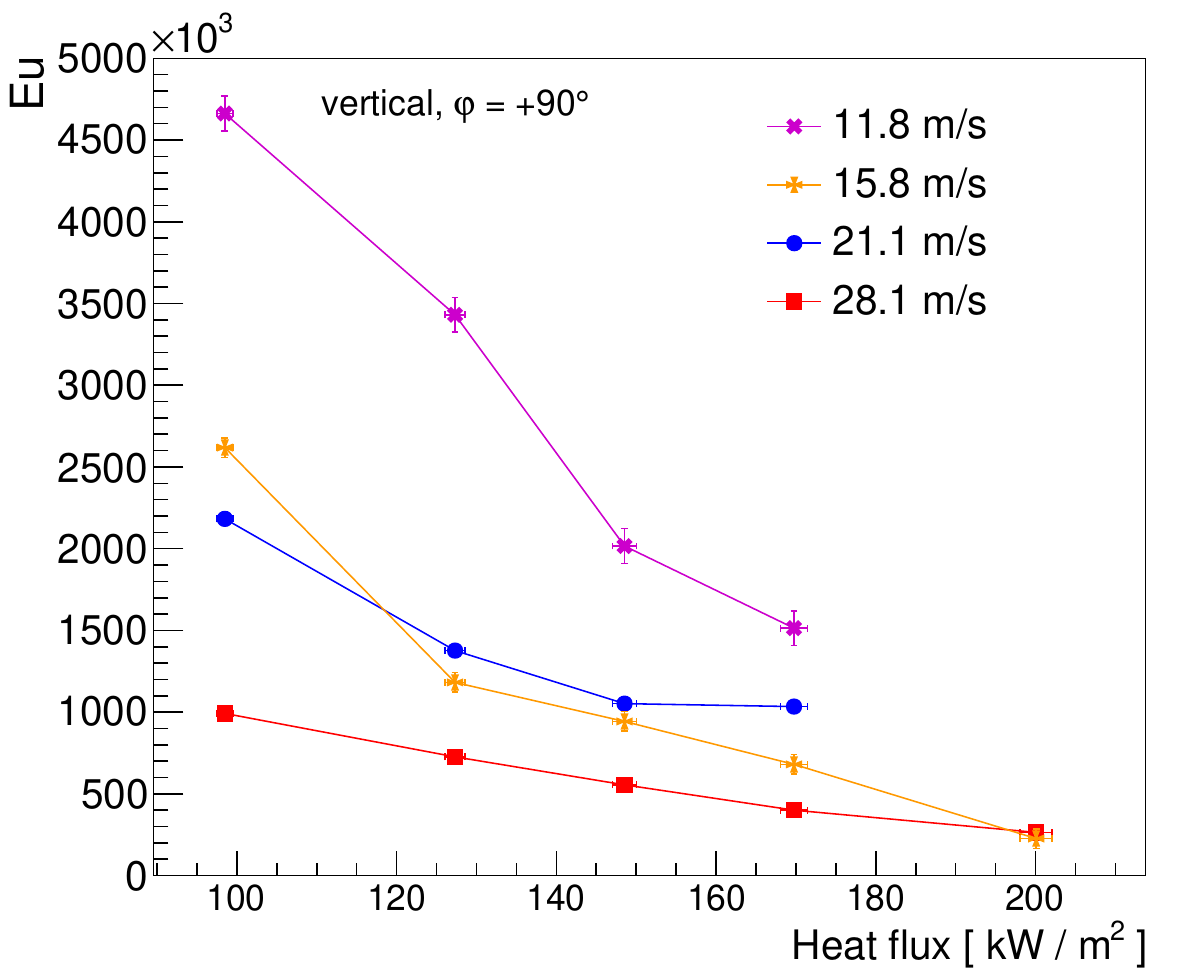}
     \label{subfig:Eu_heatflux_phi90}
  }   
  \caption{\label{fig:deltaP_heatflux}Observed pressure drop \(\Delta p\) and Euler numbers, Eu,
    as a function
    of the heat flux for different flow speeds \(v\) realised by different
    measurement setups with respect to the inner diameter of the pipes and the mass-flow
    rate, see Table~\ref{tab:velocities}.
    Horizontal flow (\(\varphi=\ang{0}\)) is shown in \protect\subref{subfig:deltaP_heatflux_phi00} and
    \protect\subref{subfig:Eu_heatflux_phi00},
    upward vertical flow (\(\varphi=+\ang{90}\)) in \protect\subref{subfig:deltaP_heatflux_phi90} and
    \protect\subref{subfig:Eu_heatflux_phi90}.
    }
\end{figure*}
In the experimental setup of this study,
the difference between the pressure before the heater and after the heater is defined as
the pressure drop \(\Delta p\).
The observed \(\Delta p\) significantly depends on properties of the setup and the CO$_2$ flow,
such as the diameter and the orientation of the pipe, the mass-flow rate, and the heat flux. 
The pressure measurements are used to determine the pressure drop as a function of
the applied heat flux. The \(\Delta p\) of a specific configuration is taken as the average
of 50 measurements across a time period of five minutes. 
Figures~\ref{fig:deltaP_heatflux}\subref{subfig:deltaP_heatflux_phi00}
and~\ref{fig:deltaP_heatflux}\subref{subfig:deltaP_heatflux_phi90} show \(\Delta p\)
observed for different flow speeds \(v\) as obtained from the different measurement setups. 
The uncertainties on the average \(\Delta p\) are displayed as error bars of the markers.
Horizontal flow (\(\varphi=\ang{0}\)) is shown in
Figure~\ref{fig:deltaP_heatflux}\subref{subfig:deltaP_heatflux_phi00},
vertical flow (\(\varphi=+\ang{90}\)) in
Figure~\ref{fig:deltaP_heatflux}\subref{subfig:deltaP_heatflux_phi90}.

For horizontal flow \(\Delta p\) increases with the heat flux. However, the relation is not
entirely linear. Larger flow speeds tend to result in a larger \(\Delta p\).
Increased turbulence with higher shear stresses and wall friction is the outcome of
increasing flow speed, leading to a frictional pressure reduction.
For annular flow \(\Delta p\) reaches a maximum, commonly identified as the
pressure-drop peak, which occurs directly before the CO$_2$ is entirely evaporated.

In the case of upward vertical flow (\(\varphi=+\ang{90}\)),
see Figure~\ref{fig:deltaP_heatflux}\subref{subfig:deltaP_heatflux_phi90},
the opposite trend to horizontal flow is observed:
\(\Delta p\) decreases with the heat flux.
The static pressure component contributes to the total pressure drop, leading to a drop
in pressure for flow in the upward direction.
At higher heat fluxes, the influence of the static pressure drop becomes less important
as the impact of friction becomes predominant.

A more explicit pattern arises among the setups implementing different flow speeds \(v\)
if the observed \(\Delta p\) is related to the kinetic energy of the fluid elements.
This relation is achieved by computing the dimensionless Euler number, Eu, which is defined by
\begin{equation}
\mathrm{Eu}=\frac{\Delta p}{\rho v^2}.
\end{equation}  
Figures~\ref{fig:deltaP_heatflux}\subref{subfig:Eu_heatflux_phi00} 
and~\ref{fig:deltaP_heatflux}\subref{subfig:Eu_heatflux_phi90} show the Euler numbers
as a function of the heat flux for different flow speeds \(v\) of horizontal flow
(\(\varphi=\ang{0}\)) and upward vertical flow (\(\varphi=+\ang{90}\)), respectively.
Flow with the lowest flow speed of \(v=\SI{11.8}{\metre\per\second}\) exhibits the largest Euler
numbers, while flow with the highest flow speed of \(v=\SI{28.1}{\metre\per\second}\)
features the lowest Euler numbers. In both cases, the Euler numbers increase with the
heat flux for horizontal flow and decrease for upward vertical flow.
The curves of configurations with intermediate flow speeds of
\(v=\SI{15.8}{\metre\per\second}\) and \(v=\SI{21.1}{\metre\per\second}\) are close
to each other and cross once. They do not fit in the pattern of lower speeds corresponding to
higher Euler numbers.

\section{Summary and conclusion}
The presented study focused on the investigation of two-phase flow of carbon dioxide (CO$_2$),
as used for
cooling detectors in particle physics. A test setup was developed and built to investigate flow
of CO$_2$ in both vertical (upward) and horizontal orientations.
The subcooled CO$_2$ was provided by a cooling plant which works with the I-2PACL system.
The CO$_2$ had a temperature of \(T=\SI{-15}{\degreeCelsius}\) and was kept at a pressure of
approximately \SI{23}{\bar}.
Two-phase flow patterns occuring in the setup after heating the CO$_2$ were recorded with a
high-speed camera. Dedicated sensors were used to measure the temperature and the pressure before and
after heating the CO$_2$.
The cooling-pipe prototypes used for the measurements were 3D printed with two different inner
diameters, \SI{3}{\milli\metre} and \SI{4}{\milli\metre}. 
The measurements were carried out with different operational parameters: Five different heat fluxes 
and four different flow speeds were considered. 

As a result, several different flow regimes were observed and classified. Stratified, wavy and slug
flow were the predominant patterns found for horizontal flow, while upward vertical flow was mainly
found to be slug or churn.

The void fraction of the different measurement setups was determined by searching for the interface
between the gaseous and the liquid phase in the recorded images. Void-fraction distributions were formed
and the average void fraction \(\bar{\alpha}\) was computed.
For horizontal flow significantly higher void fractions are found than for vertical flow.
The difference is understood as an effect of gravitation clearly separating gaseous and liquid
components into distinct phases in the case of horizontal flow, while tiny bubbles remain
dispersed in the liquid for vertical flow, leading to increased frictional forces.  

Based on the determination of \(\bar{\alpha}\) two-phase flow-pattern maps were created
for horizontal and vertical flow, following
Ref.~\cite{cheng2008}. The different measurements are marked in the plane of mass velocity \(j\) versus
vapour quality \(x\), and regions of different flow patterns are identified. Transition regions from 
one flow regime to another one are indicated. In addition to the difference in \(\bar{\alpha}\) and the
corresponding difference in \(x\) for
horizontal and vertical flow, the pattern maps also shows a less strong, but clear dependence of
\(x\) from the flow speed.

In addition, the pressure drop after heating the CO$_2$ is measured and the corresponding Euler
number is computed. While the pressure drop increases with the heat flux for horizontal flow due
to friction losses, the pressure drop reduces with the heat flux in case of upward vertical flow,
since static pressure is important in this case.

The observed differences of the studies of vertical and horizontal flow of two-phase CO$_2$
emphasize the importance of studying both separately. For detector cooling the horizontal flow
is of particular importance. Future studies could involve realistic pipes as used in the
detector and silicon-based heater elements mimicking the detector modules.

\printbibliography

\end{document}